\documentclass[iop]{emulateapj}

\shorttitle{On the Origin of Mass Segregation in NGC\,3603}
\shortauthors{Pang et al.}

\begin{document}


\title{On the Origin of Mass Segregation in NGC\,3603}


\author{Xiaoying Pang\altaffilmark{1}, Eva. K.\ Grebel}

\affil{Astronomisches Rechen-Institut, Zentrum f\"ur Astronomie der
          Universit\"at Heidelberg, M\"onchhofstr.\ 12--14, 69120
              Heidelberg, Germany}
 \email{xiaoying@ari.uni-heidelberg.de}


\author{Richard J.\ Allison\altaffilmark{2}}
\affil{Institut f\"ur theoretische Astrophysik,  Zentrum f\"ur
Astronomie der Universit\"at Heidelberg, Albert-Ueberle-Str.\ 2,
69120 Heidelberg, Germany}

\author{Simon P.\ Goodwin}
\affil{Department of Physics and Astronomy, University of Sheffield,
Sheffield S3 7RH, UK}

\author{Martin Altmann}
\affil{Astronomisches Rechen-Institut, Zentrum f\"ur Astronomie der
          Universit\"at Heidelberg, M\"onchhofstr.\ 12--14, 69120
              Heidelberg, Germany}

\author{Daniel Harbeck}
\affil{WIYN Observatory, 950 N.\ Cherry Ave, Tucson AZ 85719, USA}

\author{Anthony F.J.\ Moffat}
\affil{D\'epartement de Physique, Universit\'e de Montr\'eal,
              C.P.\ 6128, Succ.\ Centre-Ville, Montr\'eal, QC H3C 3J7,
              and Centre de Recherche en Astrophysique du Qu\'ebec, Canada}

\and
\author{Laurent Drissen}
\affil{D\'epartement de physique, de g\'enie physique et d'optique, Universit\'e Laval, Qu\'ebec QC G1V 0A6
and Centre de Recherche en Astrophysique du Qu\'ebec, Canada}

\altaffiltext{1}{Fellow of the International
          Max-Planck Research School for Astronomy and Cosmic
          Physics at the University of Heidelberg and of the
          Heidelberg Graduate School for Fundamental Physics}
\altaffiltext{2}{Alexander von Humboldt Research Fellow}


\begin{abstract}

We present deep Hubble Space Telescope/Wide Field and Planetary
Camera 2 photometry of the young HD\,97950 star cluster in the giant
H\,{\sc ii} region NGC\,3603.  The data were obtained in 1997 and
2007 permitting us to derive membership based on proper motions of
the stars. Our data are consistent with an age of 1~Myr for the
HD\,97950 cluster. A possible age spread, if present in the cluster,
appears to be small. The global slope of the
incompleteness-corrected mass function for member stars within
60$''$ is $\rm \Gamma=-0.88\pm0.15$, which is flatter than the value
of a Salpeter slope of $-1.35$. The radially varying mass function
shows pronounced mass segregation ranging from slopes of $-0.26 \pm
0.32$ in the inner $5''$ to $-0.94\pm 0.36$ in the outermost annulus
($40''$ -- $60''$). Stars more massive than 50~M$_{\odot}$ are found
only in the cluster center. The $\Lambda$ minimum spanning tree
technique confirms significant mass segregation down to
30~M$_{\odot}$. The dependence of $\Lambda$ on mass, i.e., that
high-mass stars are more segregated than low mass stars, and the
(weak) dependence of the velocity dispersion on stellar mass might
imply that the mass segregation is dynamical in origin. While
primordial segregation cannot be excluded, the properties of the
mass segregation indicate that dynamical mass segregation may have
been the dominant process for segregation of high-mass stars.

\end{abstract}



\section{Introduction}
The compact HD\,97950 cluster in the luminous giant H\,{\sc ii}
region NGC 3603 is one of the most massive young star clusters in
the Milky Way.  As the closest and densest starburst cluster
accessible at optical wavelengths, it has been subject to many
studies during the past few decades.  The cluster contains three
Wolf-Rayet stars and up to 50 O-type stars (Drissen et al. 1995).
Its total mass is estimated to be $\sim 10^4$~M$_{\odot}$ (Harayama
et al.\ 2008) with an upper dynamical mass limit of $17600 \pm
3800$~M$_{\odot}$ (Rochau et al.\ 2010).  The Wolf-Rayet stars show
characteristics of WN6 stars, but also have Balmer absorption lines
(Drissen et al.\ 1995), suggesting that these stars are actually
core hydrogen-burning rather than evolved stars (Conti et al.\ 1995;
de Koter, Heap, \& Hubeny 1997). Two of these three WR stars are
very close binaries (Schnurr et al.\ 2008).

 Based on stellar spectral
types, Melena et al.\ (2008) argue that the most massive stars in
the HD\,97950 cluster are coeval with ages of 1--2 Myr, while less
massive stars (20--40 M$_{\odot}$) show a somewhat larger age spread
of up to 4 Myr. Recent photometric studies have arrived at a range
of ages ranging from an essentially single-burst population of 1 Myr
(e.g., Sung \& Bessell 2004; Stolte et al.\ 2004; Kudryavtseva et
al.\ 2012) to 2 -- 3 Myr (Eisenhauer et al.\ 1998; Harayama et al.\
2008). The stars in the cluster outskirts may be slightly older
($\sim 5$ Myr according to Sung \& Bessell 2004), as is also
suggested by spectroscopic studies of the late O and early B-type
supergiants outside the core of the HD\,97950 cluster (Melena et
al.\ 2008). These evolved supergiants are probably not physically
connected with HD\,97950 owing to their advanced evolutionary state
(e.g., Sher\,25, see Brandner et al.\ 1997a, b) and higher age
(Crowther et al.\ 2008; Melena et al.\ 2008). They may even indicate
the occurrence of multiple episodes and possibly sequential star
formation NGC\,3603 (e.g., Moffat 1983; Melnick et al. 1989; De Pree
et al.\ 1999; Tapia et al. 2001). Beccari et al.\ (2010) suggest an
extended star formation episode of up to 10 -- 20 Myr as indicated
by an apparent age spread in pre-main-sequence stars in NGC\,3603.

Despite its young age, the HD\,97950 cluster shows pronounced mass
segregation (e.g., Sung \& Bessell 2004; Grebel \& Gallagher 2004).
Mass segregation is often observed in young star clusters (e.g., in
the ONC, Hillenbrand \& Hartmann 1998; Arches, Stolte et al.\ 2002;
NGC\,6611, Bonatto, Santos, \& Bica 2006; NGC\,2244 and NGC\,6530,
Chen et al.\ 2007; Schilbach et al.\ 2006), but the origin of mass
segregation is still unclear.  Bonnell \& Davies (1998) argue that
clusters cannot dynamically segregate in only a few Myr and so mass
segregation in young clusters must be primordial.

Whether mass segregation is primordial or dynamical is an important
constraint on theories of massive star formation and cluster
formation and evolution.  The competitive accretion theory (Bonnell
et al.\ 2001; Bonnell \& Bate 2006) suggests that protostars in the
dense central regions of a young star cluster can accrete more
material than those in the outskirts and that therefore primordial
mass segregation would be a natural outcome of massive star
formation. However, if mass segregation can occur dynamically on a
very short timescale then massive star formation can occur anywhere
in a cluster, possibly monolithically (e.g., Krumholz et al.\ 2009).

McMillan et al.\ (2007) show that young mass-segregated clusters may
be the result of mergers between small clumps that are
mass-segregated by either primordial or dynamical means.  Allison et
al.\ (2009a, 2010) suggest that observations support that clusters
form with initial substructure, and show that for clusters with
initially cool (subvirial) and clumpy distributions dynamical mass
segregation can occur very rapidly in the cluster's core after it
has collapsed ($\sim$ 0.5 -- 1 Myr).  In contrast to smooth,
subvirial clusters, clumpy clusters collapse to much higher
densities, enabling fast dynamical segregation.

The initial conditions of star clusters and the origin of mass
segregation place important constraints on theories of massive star formation and cluster evolution.
In the case of dynamical segregation,
mass segregation is expected to be
observable down to some ``limiting mass'' that is proportional to
the dynamical timescale (Allison et al.\ 2009a, 2010).
  This appears to be the case in, e.g.,
the Orion Nebula Cluster (ONC, $\sim 1$ Myr), which was found to be
mass-segregated down to 5~M$_{\odot}$ using the minimum spanning tree
(MST) method (Allison et al.\ 2009b).  Discussing the different
methods commonly used to evaluate mass segregation in star clusters,
Olczak et al.\ (2011) argue that the MST method is superior to the
other methods since it does not make assumptions about symmetry or the
location of the center of the distribution nor is it affected by
uncertainties introduced by binning (see also Allison et al.\ 2009b;
K\"{u}pper et al.\ 2011).

Here we analyze Hubble Space Telescope (HST) observations of the
massive HD\,97950 cluster in NGC\,3603 obtained with the Wide-Field
Planetary Camera 2 (WFPC2).  In Section 2 we summarize the
observations and data reduction. In Section 3 we discuss the color-magnitude
diagram of the HD\,97950 cluster.  In Section 4 we infer the present-day
mass function and discuss evidence for mass segregation based on the
traditional mass function analysis in concentric annuli. Afterwards,
we refine and quantify the mass segregation using an MST analysis.
In Section 5,
we investigate the origin of the mass segregation in the cluster with
kinematic data (tangential velocity and velocity dispersion).
 We argue that dynamical processes are the dominant mechanism for the
mass segregation in the cluster in Section 6.
We present our conclusions and summary in Section 7.

\section{Observations and Data Reduction}

For our analysis of the HD\,97950 cluster in NGC\,3603 we used deep
imaging data obtained with HST/WFPC2.  The first observations were
carried out in 1997 July (program GO 6763, PI:
Drissen).  The Planetary Camera (PC) chip was centered on the cluster.
  We obtained shallow, intermediate, and
long exposures ranging from fractions of a second to 20 -- 30~s in
the {\it F547M} and {\it F814W} filters, respectively. Details are
given in the exposure time log in Table 1. Earlier results from
analyses of these data were presented by Sung \& Bessell (2004) and
by Grebel \& Gallagher (2004).

The second data set was obtained in 2007 September (program GO 11193,
PI: Brandner). The longest
exposures lasted 100~s ({\it F555W}) and 160~s ({\it
F814W}), considerably longer than in 1997 (Table 1).
The ten-year epoch difference between the
first and the second data set permits us to infer cluster membership
using proper motions.  Preliminary results of this analysis were
presented by Pang et al.\ (2010). Rochau et al.\ (2010) published a
proper motion study of the same dataset.

Both data sets were reduced using {\em HSTphot} (Dolphin 2000, 2005),
a program developed for crowded-field stellar photometry of WFPC2
data. The shifts between the dithered images were determined
following Koekemoer et al.\ (2002).

Stars at cluster-centric distances of $20''-60''$ are located on the
three Wide Field Camera (WFC)
chips in our data.  While both in 1997 and in 2007 the PC chip was
centered on the HD\,97950 cluster, the two pointings are rotated by
$51.4\degr$ with respect to each other.  Thus the WFC chips only have
 13\% overlap, whereas the common area covered by the
PC chip exposures from the two epochs amounts to $\sim 90$\%.

We only use images obtained in the filters either common to both
datasets ({\it F814W}) or at comparable wavebands ({\it F547M} in
the 1997 dataset and {\it F555W} in the 2007 dataset). Conveniently,
{\it HSTphot} transforms magnitudes in these filters into the {\it
V} and {\em I} bands in the standard Johnson-Cousins system.
 We found 571 common
stars on the PC and WFC chips within the cluster radius of
$\sim60''$
 (Sung \& Bessell 2004) observed in both epochs. The magnitudes of
 the common stars are taken from the 1997 photometry. The position- and
magnitude-dependent incompleteness in the detection of point sources
was assessed through artificial star experiments.

Proper motions were derived using common stars observed in the same
filter during the two epochs in order to select likely cluster
members and to weed out field stars. The membership of stars in the
cluster is determined by fitting a two-Gaussian model to the proper
motion distribution. We select only stars with membership
probabilities larger than 0.7 to be cluster members (Jones \& Walker
1988), which are retained in the subsequent analysis ( see online
Table 2-4). Fifty-nine stars on the PC and WFC chips (10\% of the
total number of common stars) were thus eliminated as foreground
stars.

Because of the decreasing stellar density with
increasing cluster radius the fractional foreground contamination
increases with radius as well. Because of the high extinction in the
NGC\,3603 giant H\,{\sc ii} region, we may assume that background
stars are effectively obscured and do not significantly contribute
to our measurements.  However, giant H\,{\sc ii} regions often show
a complex age structure (e.g., Grebel \& Chu 2000), and Beccari et
al.\ (2010) found older PMS stars in a wide area around the cluster.
Hence we cannot exclude the presence of older stars belonging to
NGC\,3603 that would be difficult to disentangle from younger
cluster stars at the same distance via proper motions.

\section{Color-magnitude Diagrams and Age}

Figure 1 shows the color-magnitude diagrams (CMDs) of the HD\,97950
cluster including all common stars measured in the two epochs in the cluster core (PC: 20$''$, left panel)
 and within the cluster radius ($\sim 60''$, Sung \&
Bessell 2004; PC \& WFCs, right panel).
 The CMDs show a steep main sequence (MS) on the PC and a broader MS
(at the faint end) when the WFCs are included. The contamination of foreground stars
is more severe for the WFCs (grey dots) since they cover a larger area. There is a
broad region of redder pre-main-sequence (PMS) stars and a
wide transition region between the MS and the PMS in both CMDs.
 Like Harayama et al.\ (2008), we do not see clear evidence of a sequence of
equal-mass binaries as earlier suggested by Stolte et al.\ (2004).

 Figure 5 in Sung \& Bessell (2004) shows that $E(B-V)$ stays
unchanged with 1.25\,mag within 30$''$ (see also Moffat 1983).
We adopt a reddening law of $E(V-I)/E(B-V)=1.45\pm0.05$ and $E(B-V) = 1.25$
 from Sung \& Bessell (2004), and assume a uniform extinction of $A_V=4.44\pm0.15$ ($R_V=
3.55$) throughout the region within $r\le60''$. The reddening
corrected MS on the PC aligns with that on the WFCs (right panel of
Figure 1), indicating that our adoption is reasonable.

In order to derive
stellar masses along the main sequence, we use the isochrone models of
Lejeune \& Schaerer (2001). These isochrones extend to masses above
100 M$_\odot$, appropriate for the HD\,97950 cluster (see Schnurr et
al.\ 2008; Crowther et al.\ 2010). For the PMS stars on WFPC2 images, for which
the mass goes down to 0.8\,$\rm M_\odot$ (Drissen 1999),
we use Siess et al.\ (2000) isochrones,
which cover a larger mass range ($\rm 0.1-7.0\,M_\odot$) than other PMS
isochrones. We adopt a distance of $d=6.9\pm0.6$\,kpc from Sung \&
Bessell (2004) and solar metallicity for the HD\,97950 cluster (see
Hendry et al.\ 2008) throughout this paper.

The MS of HD\,97950 is well-represented by a Lejeune \& Schaerer 1 Myr
isochrone (Figure 1).  Slightly older MS isochrones also provide a
good fit, in agreement with spectroscopic age estimates
for the massive stars.
The Siess isochrones
along the PMS locus indicate an age spread of up to 3 Myr.
 The color uncertainties of the bulk of our PMS stars (fainter than $V=19$) are still smaller than the width of the color
distribution in that area,  which may be in part caused by differential reddening (Pang et al.\ 2011).
In the ``turn-on'' region where PMS stars
join the MS we find a broad range of luminosities ($16 \lesssim V
\lesssim 19$), which either again indicates an age spread or is due to the presence of
 (MS) stars with surviving circumstellar disks (Stolte et al.\ 2004).

Non-accreting isochrones, e.g., the Siess isochrones, tend to
 overestimate the stellar ages for stars whose effective temperature is above 3500\,K (Hosokawa et al.\ 2011).
Baraffe et al.\ (2009)
suggest that the apparent spread of the PMS stars in the Hertzsprung-Russell
 diagram at ages of a few Myr can be plausibly attributed
to a spread in the stellar radius and a different episodic accretion history, instead of
 an age range as inferred from non-accreting stellar evolutionary models (e.g., Siess 2000).

The recent study of massive MS stars in the HD\,97950 cluster by
Kudryavtseva et al.\ (2012) finds that the age spread is as small as
0.1\,Myr. A few low-mass MS stars at $V > 20$ (within $r>20''$;
right panel in Figure 1) are below the region where most of the
cluster MS stars are located (see also Grebel \& Gallagher 2004).
Considering their small proper motions, these faint MS stars are
consistent with being cluster members, which would corroborate an
age spread in the cluster as suggested in a number of earlier
studies (see Section 1). Alternatively, they might be stars from
earlier star formation in the wider NGC\,3603 H\,{\sc ii} region
that we observe superimposed at the cluster's location. That would
be consistent with the much more widely distributed population of
older PMS stars around the HD\,97950 cluster described by Beccari et
al.\ (2010).

Considering the above findings and deliberations, we excluded the MS
stars at $V > 20$ from the age determination for the HD\,97950
cluster.  We conclude that an age spread (if any) in the HD\,97950
cluster must be small.  We therefore adopt an age of 1~Myr for the
cluster.

\section{The Mass Function and Mass Segregation}



\subsection{The qualitative approach:  Mass function determination
in cluster-centric annuli}

In order to derive the mass function of the HD\,97950 cluster, we
count stars in absolute $V$-magnitude bins spaced such that they cover
mass bins with a logarithmic size of 0.2.
Using the same procedure as
Grebel \& Chu (2000), we find the absolute magnitudes corresponding to
mass bins along the earlier described isochrones assuming an age of 1
Myr.  We applied a color cut of $(V-I) = 2.4$  to separate MS and PMS
stars.

Since the crowding is severe in the central region of the HD\,97950
cluster, we corrected the count rates for incompleteness depending
on their positions and magnitudes. We display the completeness
dependence on stellar mass for the PC chip in
 Figure 2.  In the outer annulus ($r>15''$), stars above
1.5~M$_\odot$ are more than 50\% complete. As the
crowding becomes stronger towards the cluster
center, in the region within $r<5''$
only stars more massive than $\rm 4\,M_\odot$ are $>50\%$ complete.
 Therefore, many faint stars in the central region remain undetected.
 A completeness test is also run for
median and deep exposure images of WFC chips in which crowding
effects might intervene. However, since the stars on the WFC chips
($\sim60''$) are quite far from the cluster core, they are not
significantly affected by crowding. Their completeness fraction does
not depend on the cluster-centric distance (see Figure 3).

The three luminous
Wolf-Rayet stars near the center of the HD~97950 cluster are saturated
in our WFPC2 photometry.  Therefore they were added by hand to the
highest mass bin using the masses and magnitudes of Schnurr et
al.\ (2008). Since there are very few MS stars fainter than $V=20$\,mag
in the core of the cluster (left panel in Figure 1),
 the presence of a small number of stars along the lower MS ($V>20$\,mag) at
 larger cluster-centric distances (right panel in Figure 1)
may be attributed to earlier generations (Section 3).
Consequently, we exclude stars with $V>20$\,mag and $(V-I)<2.4$ from our mass function derivation.

Even though the overlap between the WFC chip exposures obtained in
1997 and 2007 is small, we can attempt to increase the area
available for analysis by also including those WFC stars in regions
that do not overlap. This greatly reduces the corrections for
missing area. While this will permit us to consider the mass
function within the entire cluster radius ($\sim 60''$; Sung \&
Bessell 2004), it also requires statistical field star subtraction.
This approach is viable for a classical mass function analysis in
which we consider the mass function within different cluster-centric
annuli, but the subsequent MST analysis is necessarily limited to
the inner $20''$ covered by the PC chip, since the MST method
requires a contiguous area.

We count the total number of incompleteness-corrected,
proper-motion-selected foreground stars in the PC and
the WFC chips in each magnitude bin considered.
Assuming that foreground stars are essentially homogeneously
distributed across the entire area covered by the WFPC2
exposures, this approach provides us with the best possible
statistics for foreground stars.
We obtain the number of foreground stars per magnitude bin
and per unit area.  In order to correct for field star
contamination, we then only need to subtract these numbers
after scaling them by the area
actually considered within a given annulus.

Fitting the corrected number counts of all probable MS and PMS stars
within a mass range of $\sim 1 - 100$~M$_\odot$ results in a mass
function slope of $\rm\Gamma=-0.82\pm0.20$ for the PC chip. Our
result is in agreement with the earlier WFPC2 study of Sung \&
Bessell (2004) within error, who obtained $\rm \Gamma=-0.9\pm0.1$
for stars on the PC chip. Combining the corrected number counts of
all stars within 60$''$, the resulting slope of the global mass
function is $\rm \Gamma=-0.88\pm0.15$ ($\rm log(mass/M_\odot)>0.6$),
which is
 flatter than a Salpeter slope of $-1.35$.

In Figure 4 we also show the
mass function of the HD\,97950 cluster in different concentric
annuli out to $60''$. Two effects stand out:
(1) the slope of the mass function
increases with radius, and (2) the more massive stars are
concentrated in the center and are missing at larger radii.

Our photometric mass function is affected by the uncertainties in
the isochrone models used to derive the masses of cluster member
stars. One such uncertainty is the unknown amount of stellar
rotation that can affect the colors and magnitudes of stars (Grebel
et al.\ 1996). The stellar evolution models with rotation
(Ekstr{\"o}m et al.\ 2012) generate a slightly narrower MS width
than non-rotating models (e.g., Schaller et al.\ 1992), and predict
larger final masses at the end of evolution for stars with initial
masses in the range of 45-100\,$\rm M_\odot$. Thus a flatter slope
of mass function will result.

 Another uncertainty is unrecognized
binarity.  Here we implicitly make the simplified assumption that we
are dealing with non-rotating, single stars as discussed in Section
3. Also, we neglect a possible age spread in the cluster, but
emphasize that a small spread such as the spectroscopically inferred
age spread of 1 -- 2 Myr for massive MS stars does not affect the
photometrically estimated masses significantly.  Moreover, the above
analysis of mass segregation in HD\,97950 is sensitive to the
determination of the position of the cluster center, and the number
and size of the radial bins used (e.g., Gouliermis et al.\ 2004).

\subsection{The quantitative approach:  Mass segregation determination
via the minimum spanning tree}

In order to quantify the mass segregation,
 we apply the $\Lambda$-method (Allison et al.\ 2009b;
Parker et al.\ 2011) to the MS members ($\rm > 3.5\,M_\odot$) on the PC
chip. We only consider MS stars
 since lower-mass stars (primarily PMS stars) are incomplete
 in the center due to crowding effects.
  We take a subset of $n$ stars of similar mass (the 1st to $n$th
most massive stars, or the $(n+1)$th to $2n$th most massive stars
for example) and find the length of the MST that connects those
stars with the shortest path without closed loops. We then take a
large number of random sets of $n$ stars of {\em any} mass and
obtain the median and the 1/6th and 5/6th percentiles to obtain a
(possibly asymmetric) $1 \sigma$ error (the vertical bar in Figure
5). A subset is mass-segregated if $\Lambda$ (the ratio of the MST
length of random stars over massive stars) is larger than unity,
i.e., the stars in that subset are more concentrated in their
distribution than a random sample (see Allison et al.\ 2009b;
Maschberger \& Clarke 2011; Parker et al.\ 2011).

Figure 5 shows the values of $\Lambda$ for samples of $20$ stars
moving in steps of $10$ stars (therefore every second datapoint is
uncorrelated with each other).  The
first $20$ stars all have masses $\rm > 35\,M_\odot$, and the second mass bin
is in the range $\rm 27 - 45\,M_\odot$.  The first two bins
have a $\Lambda$ significantly greater than
unity -- i.e., they are more concentrated than random
stars.  Varying the size of bins always shows a significant degree of
mass segregation for masses
$\rm > 30\,M_\odot$. Considering error bars, the degree of segregation
among stars with masses $\rm < 30\,M_\odot$ is not pronounced.

We note that two bins just below $\rm 20\,M_\odot$ also show
some evidence of mass segregation.  This is due to a close pair of
stars of very similar mass ($\rm\sim 18\,M_\odot$).

\section{Origin of the mass segregation}

We note that (a) the HD\,97950 cluster
is strongly mass-segregated
 above 30~M$_\odot$, and (b) all other masses of stars are
randomly distributed throughout the cluster (Figure 4 \& 5). As we
shall argue, this strongly suggests a dynamical origin for mass
segregation in HD\,97950. To verify this, we explore the kinematics
of the cluster via proper motions. Since the faint stars are
incomplete especially in the cluster center,
 in order not to bias our result, we only use stars that are more than 50\%
 complete, which corresponds
 to stars brighter
than $V=18$\,mag within the inner 5$''$ region and stars brighter
than $V=22$\,mag in the region $>5''$ from the cluster center.


\subsection{Tangential velocity profile}
 We convert the proper motions of stars into tangential velocities and show their distributions in Figure 6.
The vertical bar is the tangential velocity dispersion for stars in
each magnitude bin or annulus. The (mean) tangential velocity $V_t$
increases slightly from bright to faint stars (upper panel), and
from the inner to the outer part of the cluster (lower panel).
However, owing to the large scatter, the ascending trend is not
significant.

The tangential velocity dispersion for stars $\rm >30\,M_\odot$ is
$\rm 6.8\pm0.8\,km\,s^{-1}$. It does not change much for stars of
10\,M$_\odot$ ($\rm 5.9\pm0.6\,km\,s^{-1}$), but increases to $\rm
9.0\pm0.9\,km\,s^{-1}$ for stars of $\rm \sim 2.5\,M_\odot$. Since
the energy equipartition is mass-dependent (see Section 5.3),
dynamical segregation may only manifest itself among the few most
massive stars, considering the young age of the cluster.
 This might indicate that equipartition is not taking place in
 the entire cluster yet (Rochau et al.\ 2010). However, accounting for the
observational uncertainties (see Section 5.2), the dependence of
velocity dispersion on stellar mass is weak, similar to the finding
of Rochau et al.\ (2010).

\subsection{Velocity dispersion}

 We compute the observed one-dimensional dispersion of
 proper motions of member stars on the
PC chip, which centers at the cluster
and provides a more reliable velocity dispersion than the WFC
chips due to the higher spatial resolution.
  We compute the observed one-dimensional dispersion
(OD) of the proper motions of member stars:
$\sigma_{x,obs}=0.316\pm0.014$\,mas\,yr$^{-1}$ and
$\sigma_{y,obs}=0.325\pm0.014$\,mas\,yr$^{-1}$ ($x$ and $y$ are
pixel coordinates). We assume that the error of the observed
dispersion is given by the measurement uncertainty, consisting of
random errors from single epoch observations (1997 and 2007) and
centroid offsets.

To compute the positional
random errors of the observations, we divide the original single
 epoch data (1997 \& 2007) into two subsamples, respectively.
 After doing photometry on
each subsample, we find the common stars ($V<18$\,mag within 5$''$
from the cluster center and $V<22$\,mag for regions $>5''$ ) between
the two subsets of the same epoch.
The detected positions of the same star in the two subsamples
tend to be slightly offset from one another.
The standard deviation of this positional offset is the random error, which
 amounts to
$\sigma_{x,r}=0.262\pm0.007$\,mas\,yr$^{-1}$ and
$\sigma_{y,r}=0.233\pm0.008$\,mas\,yr$^{-1}$ when considering the
contributions from both observing epochs.

We evaluate the quality of the centroids of the detected stars by
comparing their positions in images obtained in the same filter and
with the same exposure time in the same epoch. The average
intra-filter offsets are $\sigma_{x,cent}\sim0.10$\,mas\,yr$^{-1}$
and $\sigma_{y,cent}\sim0.11$\,mas\,yr$^{-1}$. We subtract the
random errors and centroid offsets from the observed dispersion
 and obtain the absolute one-dimensional velocity dispersions:
$\sigma_{x}=0.146\pm0.016$~mas~yr$^{-1}$ and $\sigma_{y}=
0.198\pm0.016$~mas~yr$^{-1}$. By adopting a distance of $d=6.9$~kpc
from Sung \& Bessell (2004), the one-dimensional velocity
dispersions are $\sigma_{x,c}=4.8\pm0.5$~km~s$^{-1}$ and
$\sigma_{y,c}=6.5\pm0.5$~km~s$^{-1}$.
The uncertainty, $\sqrt{{\rm var}(\sigma_{x/y,c})}$, is computed according to Equation 12 in Pryor \& Meylan (1993):
\begin{equation}
 {\rm var}(\sigma_{x/y,c})=(\sigma_{x/y,c}^2+\sigma_{x/y,e}^2)^2/(2N\sigma_{x/y,c}^2)
\end{equation}
where $\sigma_{x/y,c}$ is the real one-dimensional dispersion and $\sigma_{x/y,e}$
the measurement uncertainty (random errors and centroid offsets).

\subsection{Dynamical mass of the cluster}

Harayama et al.\ (2008) derived the half-mass radius of NGC\,3603 to be $R_{hm} \sim 0.7$\,pc
from low-mass stars with masses of $\sim 0.5$ to $\sim2.5$~M$_\odot$.
As stars above 30\,$\rm M_\odot$ are significantly segregated in the center (Section 4), the true cluster half-mass radius
 would be smaller. We adopted a half-mass radius of 0.5\,pc (Rochau et al.\ 2010).
The dynamical mass $M_{dyn}$ of the HD\,97950 cluster can be
calculated from Spitzer's (1987) formula using the one-dimensional
dispersion we derived ($\rm 4.8-6.5\,km\,s^{-1}$):
\[
M_{dyn} \sim \eta \frac{R_{hm} \sigma_{vt}^2}{G}
\]

$\eta$ is a dimensionless parameter in the equation.  The square of
the three-dimensional velocity dispersion is three times larger than
the square of the one-dimensional dispersion, $\sigma_{vt}^2$.
Moreover, we are taking a factor of $4/3$ from the projection of
half-light radius on the sky into account.  Furthermore, there is a
factor of 5/2 from the conversion to a star cluster fitted with a
King mass profile. Thus $\eta$ is about 10.0 (see Fleck et al.\ 2006
for details). Observers usually use $\eta$=9.75 when working with
the half-mass radius $ R_{hm}$ instead of the half-light radius.

This results in a dynamical mass of $M_{dyn} \sim 1.9\pm0.6 \times
10^4$~ M$_\odot$ ($\eta=9.75$). This mass is close to the
photometric mass $M_{phot}= 1-1.6\times10^4$\,$\rm M_\odot$ derived
from observations of the stellar content (Harayama et al.\ 2008),
which might suggest NGC\,3603 is more or less virialized. Several
other young star clusters are also found to be virialized with
comparable photometric and dynamical masses, e.g., Westerlund 1:
Cottaar et al.\ (2012); R\,136, Bosch et al.\ (2009);
 ONC, Jones \& Walker (1988), Tobin et al.\ (2009).

 Compared to the previous study of Rochau et al.\ (2010) of
NGC\,3603 with the same dataset, our study differs from theirs in
the following aspects: (1) The velocity dispersion we derived is
based on member stars with a completeness of more than 50\%. These
are stars brighter than $V=18$\,mag within 5$''$ from the center and
stars brighter than $V=22$\,mag in the annulus $5''-20''$. Rochau et
al.\ (2010) computed the velocity dispersion for intermediate-mass
stars ($\rm 1.7-9.0\,M_\odot$) in the magnitude range of
$16<V<20$\,mag within 15$''$. As the photometric uncertainty
increases towards fainter stars, this might explain why Rochau et
al.\ (2010) arrived at a smaller value ($4.5\pm0.8$\,km\,s$^{-1}$)
for the velocity dispersion. (2) The sinusoidal pixel phase error
and breathing error on the pixel scale are smaller than the
astrometric uncertainty of HSTphot (0.03\,pixels). Therefore we did
not subtract these two error sources which are considered in Rochau
et al.\ (2010). Nevertheless, the dynamical mass of our study and
that of Rochau et al.\ (2010) ($\rm 17600\pm3800\,M_\odot$)
 agree with each other within the errors.

\section{Dynamical segregation in the cluster core}

Very similar patterns of mass segregation have been observed in
the two other clusters analyzed with the $\Lambda$-method: the
ONC (Allison et al.\ 2009b) and
Trumpler~14 (Sana et al.\ 2010).
Allison et al.\ (2009a) proposed a dynamical origin for mass
segregation due to two-body relaxation in a dense phase to explain the
ONC.  They proposed that the ONC underwent a short-lived dense phase
in which the two-body relaxation time was very short.

The two-body relaxation time $t_{\rm relax}$ of a system is given by
\[
t_{\rm relax} \sim \frac{N}{8 {\rm ln}N} t_{\rm cross}
\]
where $N$ is the number of stars in the cluster, and $t_{\rm cross}$ is the crossing time of the system.  Dynamical
mass segregation occurs due to the equipartition of energies in
two-body encounters.  The rate at which a star will approach equipartition depends on the mass
of that star, $M$, relative to the average mass of stars in the system,
$\langle m \rangle$.
The time to segregate $t_{\rm seg}$ down to a mass $M$ is
\begin{equation}
 t_{\rm seg} (M) \sim \frac{\langle m \rangle}{M} t_{\rm relax} =
\frac{\langle m \rangle}{M} \frac{N}{8 {\rm ln}N} t_{\rm cross}.
\end{equation}
Allison et al. (2009a) showed that for a dense phase that lasts one
crossing time the ONC should mass-segregate down to 5\,$\rm
M_\odot$, but not below, which is what is observed.

In Allison et al.'s (2009a) simulation, they find that in the core
 the segregation time is similar to the crossing time.
To extend their argument to the HD\,97950 cluster, we need to input the total stellar number $N$ and
mean stellar mass $\langle m \rangle$ into equation (2).
N$\rm \ddot{u}$rnberger et al.\ (2002) found about 10216 bona-fide stars to construct the luminosity
 function of the HD\,97950 cluster from infrared observations. We adopted their result and
assume a total number of stars, $N=10^4$,
 and a mean stellar mass of $\langle m \rangle=0.4\,M_\odot$ from Kroupa et al.'s (2002) initial mass function.
  Inserting $N=10^4$ and $\langle m \rangle=0.4$~M$_\odot$ into equation (2)
suggests that in one crossing time the HD\,97950 cluster should
mass-segregate to a mass of 30\,$\rm M_\odot$. This is exactly what
is observed in the core of the cluster. The highest-mass stars are
sinking further into the core of the cluster, and are found to be
more mass-segregated than lower-mass stars (Figure 5). Higher-mass
stars are more efficient at this process and thus mass-segregate
faster.  That the stars in the HD\,97950 cluster show the same
dependence on mass as do purely dynamical and initially
non-mass-segregated $N$-body simulations (Allison et al.\ 2009a,
2010) indicates that dynamical mass segregation may be the dominant
process for the mass segregation in the cluster. Furthermore, no
formation process (e.g., competitive accretion) is known to produce
mass segregation down to one particular mass. Recent infrared
observations by Gvaramadze et al.\ (2012) and Roman-Lopes (2012)
find a (bow-shock-producing) massive star and a massive binary in
the vincinity of NGC\,3603, which are suggested to have been ejected
from the cluster HD\,97950 via a three-body encounter. These
observations may imply that vivid dynamical evolution is already
taking place inside the cluster.


We also note that binaries may shorten the relaxation time and accelerate segregation.
But at the same time, primordial binaries increase the chance of ejections of OB stars (Portegies Zwart et al. 2010),
working against the observed mass segregation in the HD\,97950 cluster.
 As no binary sequence can be seen in Figure 1, we cannot quantify
the presence of binaries in the cluster and
their effects on the mass segregation.

\section{Summary}

We analyzed broad-band HST/WFPC2 imaging of the HD\,97950 star cluster
obtained in 1997 and in 2007.  We used the epoch difference to
establish a proper-motion-selected sample of probable cluster members.
The main results of our subsequent analysis are:

1. We find pronounced mass segregation in the cluster, as did
previous studies.  The slope of the mass function, measured within
concentric annuli around the cluster center, varies radially from
$-0.26 \pm 0.32$ within $5''$ from the center to $-0.94 \pm 0.36$ in
an annulus of $40''$ -- $60''$ around the center. Very massive stars
are only found near the cluster center and are not observed at
larger radii. The global slope of the mass function for member stars
on the PC chip is $\rm \Gamma=-0.82\pm0.20$.
 It stays almost unchanged at a value of $\rm \Gamma=-0.88\pm0.15$ for all stars ($\rm
log(mass/M_\odot)>0.6$) within the cluster radius of $\sim 60''$ (Sung \&
Bessell 2004),
 which is flatter than a Salpeter slope.


2. Using the $\Lambda$ MST method, we find the HD\,97950 cluster to
be significantly mass-segregated down to 30~M$_\odot$. The most
massive stars are the most mass-segregated ones.  A simple extension
of the Allison et al.\ (2009a) dynamical model for mass segregation
in the ONC suggests that HD\,97950 should be mass-segregated to
30\,M$_\odot$, in very good agreement with the observations.
Furthermore, we find a weak dependence of the tangential velocity
dispersion on the stellar mass. The tangential velocity dispersion
increases from $\rm 6.8\pm0.8\,km\,s^{-1}$ for stars $\rm
>30\,M_\odot$ to $\rm 9.0\pm0.9\,km\,s^{-1}$ for stars of $\rm \sim
2.5\,M_\odot$. Considering the uncertainty, this suggests that
energy equipartition does not affect the whole cluster yet, but so
far only the most massive stars of the HD\,97950 cluster.

3.\ We compute a dynamical mass of $M_{dyn} \sim (1.9\pm0.6) \times
10^4$~ M$_\odot$ for the HD\,97950 cluster in NGC\,3603, which is
close to its photometric mass (Harayama et al.\ 2008). This may
imply that the cluster is in a state of virialization, similar to
other young massive clusters.

\acknowledgments

We thank Rainer Spurzem and Siegfried
R\"oser for helpful discussions.  X.P.\ acknowledges support within
the framework of the Excellence Initiative by the German Research
Foundation (DFG) through the Heidelberg Graduate School of Fundamental
Physics (grant number GSC 129/1).  This work was partially supported
by Sonderforschungsbereich 881 ``The Milky Way System'' of the German
Research Foundation (subproject B5). LD and AFJM are grateful for
financial assistance from NSERC (Canada) and FQRNT (Quebec).

\clearpage



\begin{deluxetable}{cccccccc}
\tablecolumns{7}
\tablewidth{0pc}
\tablecaption{Exposure time log of the HST/WFPC2 observations of
the HD\,97950 cluster}
\tablehead{
\colhead{Filter} & \colhead{Shallow}  & \colhead{No.\ of} & \colhead{Median} & \colhead{No.\ of}
 & \colhead{Deep} & \colhead{No.\ of} \\
\colhead{ } & \colhead{Exposures} & \colhead{Frame} & \colhead{Exposures} & \colhead{Frame} & \colhead{Exposures} & \colhead{Frame}\\
& \colhead{[s]} & \colhead{ } & \colhead{[s]} & \colhead{ } & \colhead{[s]} & \colhead{ }}
\startdata
  &  & & 1997 & &  &    \\
\hline                        
   {\em F547M} & 1  & 3 & 10 &12 & 30 & 8 \\      
   {\em F814W} & 0.4 & 3 & 5 & 12 & 20 & 8 \\
\hline                        
       &  &  & 2007 & & &    \\
\hline                        
   {\em F555W} & 0.4 & 4 & 26 & 4 & 100 & 4 \\
   {\em F814W} &    --& -- & 18 & 4 & 160 & 4 \\
\hline
\enddata
\end{deluxetable}

\begin{figure}[h]
\includegraphics[width=8cm]{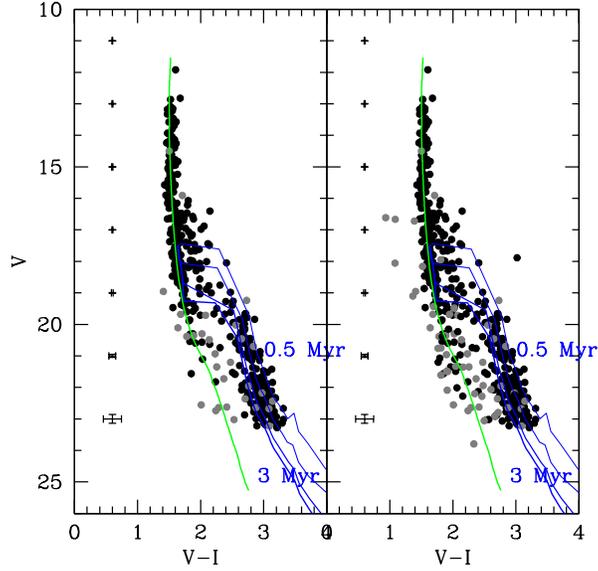}
\centering
\caption{CMDs of all common stars found in the 1997 and 2007 WFPC2 data
 (left: within 20$''$, PC; right: within 60$''$, PC \& WFC chips).
Proper-motion members of the HD\,97950 cluster are shown as black
dots. Probable non-members are indicated as grey dots.
The vertical solid green line is a 1~Myr MS isochrone for solar abundance
from Lejeune \& Schaerer (2001). The blue lines are PMS isochrones
from Siess et al.\ (2000). From right to left (increase of thickness
of the lines) isochrones for 0.5, 1,
2, 3 Myr are plotted. Representative mean errors of magnitude and color
are indicated on the left.
\label{Fig.1} }
\end{figure}

\begin{figure}
\includegraphics[width=10cm]{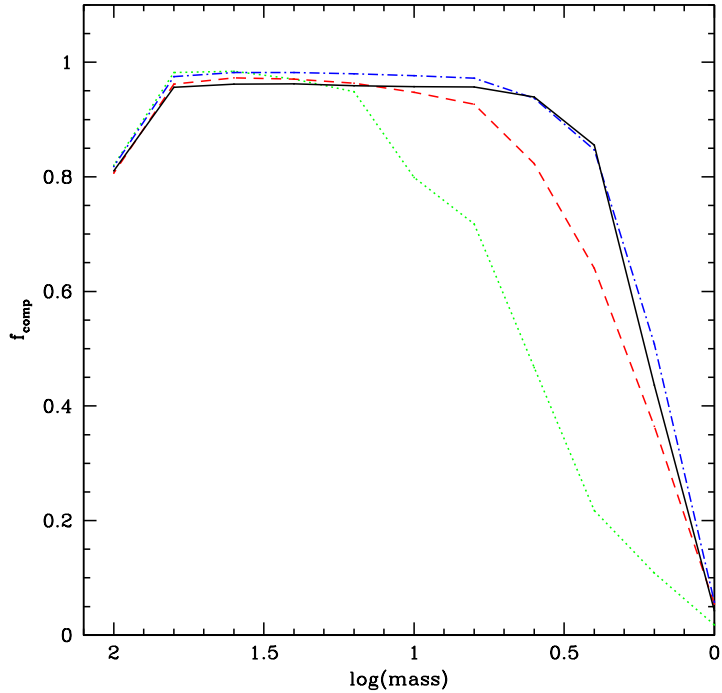}
\centering \caption[Completeness fraction as a function of the
stellar mass of the stars on the PC chip]{Completeness fraction
$f_{comp}$ as a function of the stellar mass (in solar masses) of
the stars on the PC chip.  The uncertainty of the logarithmic
stellar mass is $\pm 0.2$. The green dotted line denotes the
$f_{comp}$ distribution for a circle with a radius of $r<5''$ around
the center of the HD\,97950 cluster. The dashed line shows the
completeness fraction for the next larger annulus between $5''$ to
$10''$, the dot-dashed line indicates $f_{comp}$ for the annulus
between $ 10''$ to $15''$, and the solid line is for the region
outside $r>15''$. The drop in completeness at the high-mass end is
primarily due to saturation. Moreover, crowding effects, occasional
closeness to the chip boundaries or the location in the overlap
areas between the chips all contribute to the completeness fraction
never reaching 1. \label{Fig.2}}
\end{figure}

\begin{figure}
\includegraphics[width=10cm]{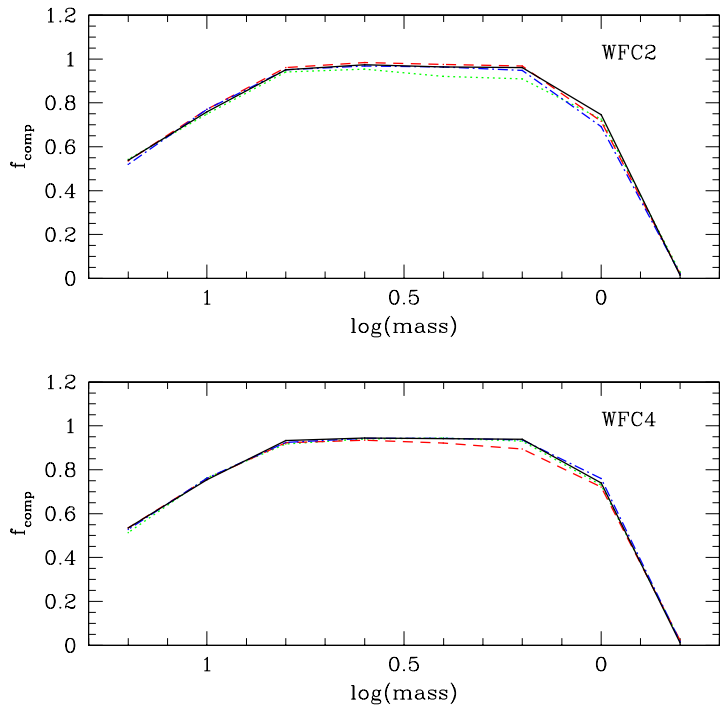}
\centering
\caption[Completeness fraction as a function of the
stellar mass of the stars on the WFC chips]{Completeness fraction $f_{comp}$
as a function of magnitude $V$ of the stars on the WFC chips.
The green dotted line denotes the $f_{comp}$ distribution for an annulus
 between 15$''$ to 25$''$ around the center of the HD\,97950
cluster. The dashed line shows the completeness fraction for the next
larger annulus between $25''$ to $35''$, the dot-dashed line indicates
$f_{comp}$ for the annulus between $ 35''$ to $45''$, and the solid
line is for the region $45''<r<60''$.
\label{Fig.3}}
\end{figure}

\begin{figure}[h]
\includegraphics[width=8cm]{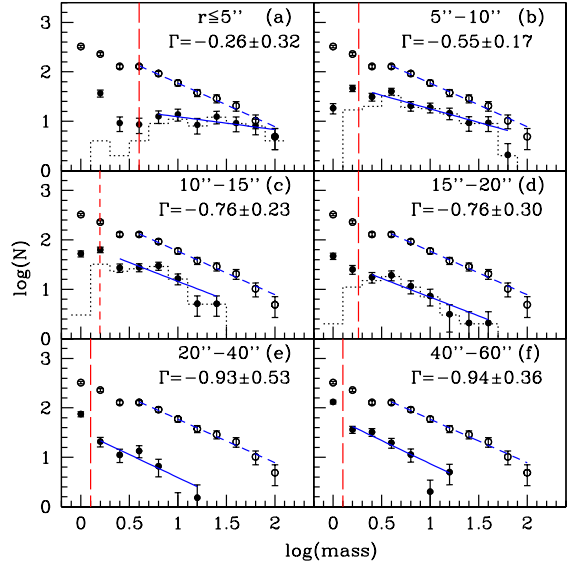}
\centering
\caption{Radial mass function of all cluster member stars on the PC
 and possible members on the WFC chips.  Panels (a) to (f) show mass functions in a sequence of
increasing annuli.  The dotted histogram is the observed mass
function for annuli within 20$''$. The filled black dots indicate
the mass function corrected for incompleteness and for foreground
contamination (panels a--f).  The vertical dashed red line indicates
a completeness limit of 50\%.  The blue solid lines show an
error-weighted linear least-squares fit of the mass function of each
annulus (black filled dots). The short-dashed blue lines are
weighted linear least-squares fits to the total mass function (open
circles) within $r\le60''$. The resulting slope of the total mass
function is $\Gamma=-0.88\pm0.15$.  The slope of the mass function
in each annulus is indicated in each panel. \label{Fig.4}}
\end{figure}

\begin{figure}[h]
\includegraphics[width=8cm,angle=-90]{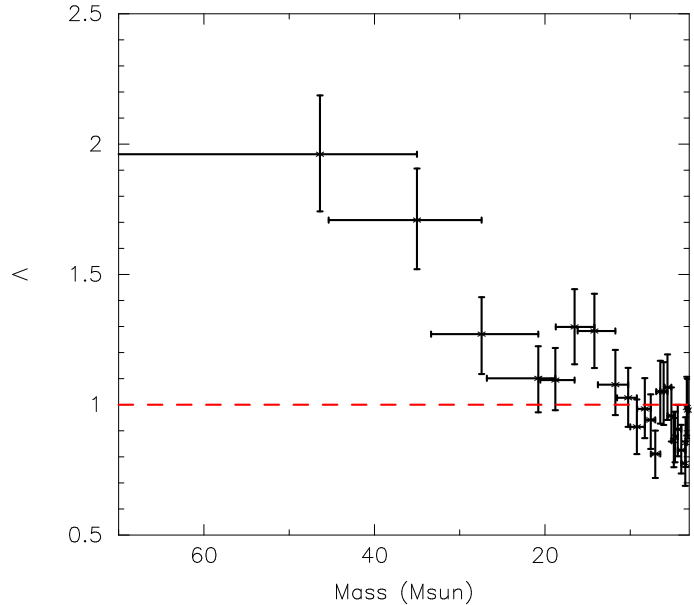}
\centering \caption{ The evolution of $\Lambda$ along stellar mass.
$\Lambda$ is the ratio of the MST length of $20$ random stars over
$20$ massive stars moving in a declining order of mass and in steps
of $10$ stars. The vertical bar is 1\,$\sigma$ error of $\Lambda$.
The dashed line indicates a $\rm \Lambda$ of unity, meaning no mass
segregation. \label{Fig.3}}
\end{figure}

\begin{figure}[h]
\includegraphics[width=10cm]{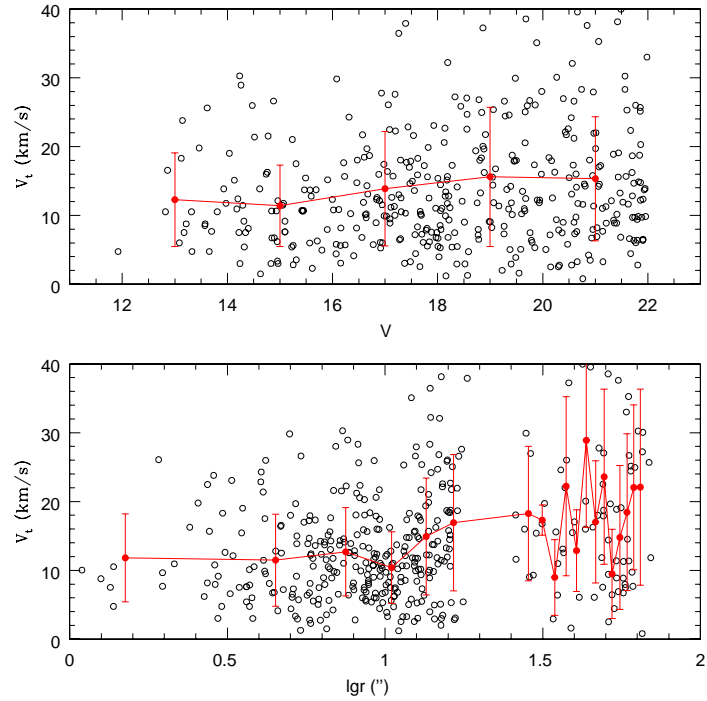}
\centering \caption[Dependence of the tangential velocity on the $V$
magnitude and the cluster-centric distance]{Dependence of the
tangential velocity on the $V$ magnitude and the cluster-centric
distance. Only stars with a completeness of more than 50\% are shown
in the plot. The filled dots show the mean velocity in each
magnitude bin (binsize = 2~mag; upper panel) and in each annulus
(annulus width $= 3''$; lower panel). The vertical bar shows the
corresponding velocity dispersion. \label{Fig.22}}
\end{figure}

\clearpage








\end{document}



\begin{deluxetable}{ccccccccccccccc}[]
\tablecolumns{8} \tablewidth{0pc} \tablecaption{The $VI$ photometry
and relative proper motions of the member stars on the PC chip of
WFPC2 (online Table 2).} \tablehead{ \colhead{No.} & \colhead{$X$} &
\colhead{$Y$} & \colhead{$V$} & \colhead{$I$} & \colhead{$r('')$} &
\colhead{$\mu_x$(mas/yr)} & \colhead{$\mu_y$(mas/yr)}} \startdata
 \hline
 1 & 430.97 & 515.83 & 15.31 & 13.79 &  4.16 & -0.32063 &  0.42928 \\
  2 & 707.94 & 502.85 & 15.24 & 13.60 & 13.74 & -0.37145 & -0.52531 \\
  3 &  79.76 & 548.09 & 15.25 & 13.70 & 16.59 & -0.06742 &  0.05419 \\
  4 & 385.40 & 354.77 & 15.10 & 13.60 &  3.63 &  0.05419 &  0.22629 \\
  5 & 444.91 & 284.16 & 15.25 & 13.71 &  6.64 &  0.14993 & -0.08465 \\
  6 & 279.54 & 568.62 & 15.45 & 13.93 &  9.17 & -0.04913 &  0.32400 \\
  7 & 557.71 & 456.41 & 15.31 & 13.77 &  6.52 & -0.06514 &  0.08690 \\
  8 & 407.94 & 298.04 & 15.43 & 13.88 &  5.92 &  0.18306 &  0.26771 \\
  9 & 516.83 & 420.49 & 15.60 & 14.01 &  4.50 & -0.25409 &  0.29719 \\
 10 & 279.54 & 568.62 & 15.42 & 13.93 &  9.17 & -0.04913 &  0.32400 \\
 11 & 456.94 & 492.47 & 15.49 & 14.01 &  3.51 &  0.12929 &  0.45399 \\
 12 & 544.61 & 364.36 & 15.47 & 13.95 &  6.43 &  0.21759 &  0.35839 \\
 13 & 635.45 & 511.55 & 15.71 & 14.15 & 10.70 & -0.19092 &  0.01979 \\
 14 & 617.10 & 474.51 & 15.57 & 13.97 &  9.37 & -0.01656 & -0.18250 \\
 15 & 529.95 & 486.75 & 15.62 & 14.07 &  5.81 & -0.03313 & -0.06219 \\
 16 & 307.54 & 535.20 & 15.57 & 14.14 &  7.18 &  0.23008 &  0.34927 \\
 17 & 621.77 & 258.26 & 16.07 & 14.27 & 12.11 &  0.14319 &  0.10486 \\
 18 & 111.78 & 470.14 & 15.69 & 14.19 & 14.28 &  0.17895 &  0.37945 \\
 19 & 401.23 & 486.53 & 15.90 & 14.45 &  2.89 &  0.21113 &  0.12059 \\
 20 & 113.38 & 486.10 & 15.80 & 14.23 & 14.33 &  0.07268 &  0.13336 \\
 21 & 460.79 & 473.05 & 16.02 & 14.53 &  2.88 & -0.31459 & -0.09939 \\
 22 & 512.66 & 392.70 & 16.41 & 14.26 &  4.57 & -0.03173 &  0.12227 \\
 23 & 327.54 & 484.04 & 16.08 & 14.58 &  4.98 & -0.86166 &  0.29901 \\
 24 & 445.04 & 601.72 & 15.91 & 14.40 &  8.16 &  0.00969 &  0.46382 \\
 25 & 284.47 & 448.25 & 16.00 & 14.49 &  6.28 & -0.31094 &  0.21296 \\
 26 & 273.67 & 439.32 & 16.21 & 14.42 &  6.72 & -0.49077 &  0.22405 \\
 27 & 444.34 & 510.54 & 16.32 & 14.81 &  4.05 & -0.63803 &  0.38001 \\
 28 &  78.40 & 300.69 & 16.16 & 14.53 & 16.70 & -0.02625 &  0.09012 \\
 29 & 391.77 & 478.92 & 16.16 & 14.68 &  2.73 &  0.16284 &  0.18951 \\
 30 & 350.35 & 341.22 & 16.07 & 14.52 &  5.03 &  0.04183 &  0.12831 \\
 31 & 243.53 & 570.81 & 16.15 & 14.65 & 10.46 & -0.03720 &  0.16790 \\
 32 & 206.53 & 509.92 & 16.12 & 14.64 & 10.51 & -0.29852 &  0.14431 \\
 33 & 464.23 & 466.28 & 16.23 & 14.73 &  2.77 & -0.22952 &  0.42044 \\
 34 & 464.23 & 238.36 & 16.34 & 14.76 &  8.89 &  0.17337 &  0.26125 \\
 35 & 286.48 & 587.12 & 16.57 & 14.70 &  9.59 & -0.10023 &  0.65642 \\
 36 & 504.90 & 503.73 & 16.24 & 14.71 &  5.32 & -0.16242 & -0.06429 \\
 37 & 489.07 & 675.78 & 16.46 & 14.68 & 11.92 &  0.34127 &  0.29508 \\
 38 & 230.44 & 136.61 & 16.60 & 14.86 & 15.90 &  0.55240 &  0.11455 \\
 39 & 572.25 & 340.07 & 16.47 & 14.78 &  8.08 & -0.14235 &  0.53303 \\
 40 & 522.21 & 564.68 & 16.46 & 14.83 &  7.94 &  0.23612 &  0.08142 \\
 41 & 210.33 & 156.48 & 16.70 & 14.97 & 15.69 &  0.51471 &  0.24840 \\
 42 & 317.97 & 375.49 & 16.66 & 14.77 &  5.21 & -0.11638 &  0.59072 \\
 43 & 502.33 & 376.68 & 16.56 & 14.98 &  4.45 &  0.15512 &  0.13856 \\
 44 & 766.53 & 443.54 & 16.67 & 15.03 & 16.00 & -0.15695 & -0.36850 \\
 45 & 401.53 & 512.21 & 17.44 & 15.32 &  4.04 & -0.69910 &  0.00449 \\
 46 & 451.09 & 399.54 & 17.06 & 15.25 &  1.91 & -0.17351 & -0.77827 \\
 47 & 505.32 & 258.49 & 16.77 & 15.14 &  8.67 & -0.03580 &  0.17239 \\
 48 & 376.85 & 433.00 & 16.64 & 15.13 &  1.97 & -0.19738 &  0.21984 \\
 49 & 339.71 & 440.82 & 16.63 & 15.10 &  3.72 & -0.28750 &  0.11273 \\
 50 & 213.40 & 397.73 & 16.88 & 14.83 &  9.55 & -0.29122 &  0.39138 \\
 51 & 250.75 & 313.46 & 16.59 & 14.94 &  9.32 &  0.06071 &  0.35881 \\
 52 & 508.77 & 415.49 & 17.03 & 15.40 &  4.15 & -0.29522 &  0.06429 \\
 53 & 553.60 & 460.42 & 16.83 & 15.19 &  6.38 & -0.00590 &  0.27332 \\
 54 & 476.66 & 441.44 & 17.09 & 15.50 &  2.74 & -0.01474 &  0.68716 \\
 55 & 458.81 & 631.43 & 16.98 & 14.97 &  9.61 &  0.32807 &  0.41946 \\
 56 & 182.45 & 455.58 & 16.64 & 15.07 & 10.97 & -0.31193 & -0.01741 \\
 57 & 574.46 & 582.99 & 16.84 & 15.18 & 10.15 & -0.37426 &  0.04296 \\
 58 & 381.50 & 256.05 & 16.89 & 15.22 &  8.02 & -0.35278 &  0.11638 \\
 59 & 359.85 & 620.92 & 17.49 & 15.35 &  9.36 &  0.45217 &  0.30771 \\
 60 & 282.83 & 503.05 & 16.65 & 15.18 &  7.20 & -0.43476 & -0.05377 \\
 61 & 390.49 & 543.12 & 17.46 & 15.56 &  5.54 & -0.19962 & -0.47954 \\
 62 & 379.32 & 246.33 & 16.95 & 15.19 &  8.48 & -0.00365 &  0.18355 \\
 63 & 486.73 & 425.43 & 16.85 & 15.24 &  3.11 & -0.02808 &  0.38450 \\
 64 & 484.66 & 300.92 & 16.94 & 15.29 &  6.50 & -0.21254 &  0.28203 \\
 65 & 435.64 & 515.56 & 17.09 & 15.55 &  4.18 &  0.10908 &  0.26897 \\
 66 & 558.53 & 473.92 & 17.09 & 15.49 &  6.78 &  0.10585 &  0.25030 \\
 67 & 244.54 & 139.89 & 17.17 & 15.48 & 15.43 &  0.32842 &  0.10809 \\
 68 & 146.55 & 393.10 & 16.92 & 15.29 & 12.63 & -0.06345 &  0.14768 \\
 69 & 540.35 & 668.07 & 17.15 & 15.28 & 12.44 & -0.43799 &  0.19681 \\
 70 & 411.99 & 403.55 & 17.19 & 15.62 &  1.09 &  0.18250 &  0.24651 \\
 71 & 519.81 & 297.07 & 17.20 & 15.35 &  7.53 & -0.07833 &  0.67607 \\
 72 & 320.04 & 447.85 & 17.05 & 15.51 &  4.67 & -0.12971 &  0.48403 \\
 73 & 299.46 & 703.65 & 17.26 & 15.60 & 13.90 &  0.86208 &  0.70668 \\
 74 & 250.97 & 496.05 & 16.99 & 15.45 &  8.38 & -0.08605 & -0.29115 \\
 75 & 150.01 & 446.89 & 17.11 & 15.59 & 12.42 & -0.32982 & -0.09012 \\
 76 & 356.35 & 271.34 & 17.13 & 15.53 &  7.69 &  0.04562 &  0.38310 \\
 77 & 572.93 & 791.49 & 17.39 & 15.63 & 18.23 & -0.84397 &  0.79428 \\
 78 & 270.57 & 441.35 & 17.54 & 15.95 &  6.87 & -0.44950 &  0.48445 \\
 79 & 473.68 & 382.05 & 17.30 & 15.69 &  3.23 &  0.10809 &  0.17056 \\
 80 & 752.85 & 244.18 & 17.19 & 15.56 & 17.49 &  0.75441 &  0.37854 \\
 81 & 314.85 & 525.04 & 17.31 & 15.66 &  6.61 & -0.27781 &  0.16228 \\
 82 & 604.62 & 451.54 & 17.30 & 15.70 &  8.61 &  0.45427 & -0.31277 \\
 83 & 568.64 & 233.36 & 17.62 & 15.64 & 11.22 & -0.22854 & -0.11273 \\
 84 & 521.79 & 476.31 & 17.52 & 15.74 &  5.25 & -0.35292 & -0.29213 \\
 85 & 555.38 & 670.06 & 17.41 & 15.60 & 12.85 & -0.53317 & -0.09518 \\
 86 & 151.26 & 451.30 & 17.66 & 15.76 & 12.38 &  0.07173 &  0.03130 \\
 87 & 451.83 & 568.40 & 17.26 & 15.70 &  6.71 & -0.27374 & -0.28890 \\
 88 & 511.70 & 159.67 & 17.68 & 15.83 & 12.98 &  0.21998 & -0.22033 \\
 89 & 604.04 & 218.05 & 17.48 & 15.84 & 12.81 &  0.00646 &  0.44711 \\
 90 & 449.27 & 215.57 & 17.39 & 15.71 &  9.79 &  0.19780 &  0.24609 \\
 91 & 670.74 & 386.80 & 17.48 & 15.85 & 11.71 & -0.09939 &  0.01334 \\
 92 & 193.08 & 584.60 & 17.37 & 15.87 & 12.70 &  0.01699 &  0.09153 \\
 93 & 671.99 & 184.11 & 17.67 & 15.76 & 16.10 &  0.32175 &  0.03629 \\
 94 & 467.53 & 605.53 & 17.62 & 15.99 &  8.54 &  0.41497 &  0.25325 \\
 95 & 557.67 & 470.34 & 17.81 & 16.13 &  6.69 &  0.36920 &  0.30266 \\
 96 & 543.90 & 442.44 & 17.80 & 16.11 &  5.79 & -0.08928 & -0.23135 \\
 97 & 517.16 & 241.73 & 17.58 & 15.97 &  9.61 & -0.02555 &  0.13343 \\
 98 & 193.34 & 554.39 & 17.62 & 16.04 & 11.94 & -0.12375 & -0.54552 \\
 99 & 409.75 & 345.83 & 17.78 & 16.12 &  3.72 &  0.07581 &  0.22812 \\
100 & 477.45 & 543.01 & 17.56 & 15.97 &  6.00 &  0.25170 & -0.33130 \\
101 & 316.49 & 413.99 & 17.57 & 15.97 &  4.76 & -0.21998 & -0.01334 \\
102 & 357.70 & 534.96 & 17.75 & 16.21 &  5.75 & -0.47322 & -0.23191 \\
103 & 391.92 & 654.89 & 17.96 & 16.13 & 10.59 &  0.33621 & -0.03173 \\
104 & 510.25 & 472.14 & 17.84 & 16.20 &  4.69 &  0.01656 &  0.38226 \\
105 & 461.91 & 527.86 & 17.73 & 16.17 &  5.07 &  0.01979 &  0.18502 \\
106 & 250.08 & 586.77 & 17.70 & 16.06 & 10.73 &  0.12003 &  0.18699 \\
107 & 497.25 & 528.50 & 17.81 & 16.16 &  5.92 & -0.16888 &  0.01067 \\
108 & 564.22 & 162.64 & 17.69 & 16.12 & 13.84 &  0.08282 & -0.29529 \\
109 & 427.06 & 127.39 & 17.82 & 16.20 & 13.75 &  0.10023 & -0.59936 \\
110 & 264.47 & 598.97 & 17.84 & 16.29 & 10.67 &  0.18811 &  0.01544 \\
111 & 353.77 & 218.74 & 17.97 & 16.32 & 10.01 &  0.13799 &  0.15547 \\
112 & 495.22 & 666.32 & 18.05 & 15.64 & 11.59 &  0.03229 &  0.25942 \\
113 & 455.06 & 364.98 & 17.93 & 16.29 &  3.27 & -0.66976 & -0.22124 \\
114 & 470.14 & 230.05 & 18.01 & 15.91 &  9.32 &  0.13252 & -0.15547 \\
115 & 179.39 & 681.67 & 18.14 & 16.21 & 16.12 &  0.48488 &  0.49583 \\
116 & 575.56 & 539.14 & 18.07 & 16.08 &  8.87 & -0.08367 & -0.22826 \\
117 & 571.41 & 188.10 & 18.22 & 16.21 & 13.00 & -0.20468 & -0.19962 \\
118 & 198.09 & 379.40 & 17.99 & 16.11 & 10.39 & -0.01607 &  0.22910 \\
119 & 603.89 & 188.19 & 18.02 & 16.34 & 13.86 & -0.08142 & -0.09019 \\
120 & 312.83 & 676.51 & 18.15 & 16.40 & 12.51 &  0.30168 &  0.03594 \\
121 & 386.64 & 426.83 & 18.07 & 16.18 &  1.50 &  0.39096 &  0.40023 \\
122 & 655.30 & 382.07 & 18.16 & 16.45 & 11.05 & -0.02583 &  0.02667 \\
123 & 562.86 & 303.39 & 18.24 & 16.58 &  8.70 & -0.33635 &  0.04913 \\
124 & 331.55 & 619.25 & 18.06 & 16.44 &  9.75 &  0.16551 &  0.28188 \\
125 & 684.28 & 368.62 & 18.18 & 16.47 & 12.48 &  0.35376 &  0.06219 \\
126 & 611.78 & 306.67 & 18.15 & 16.42 & 10.43 &  0.10725 &  0.37636 \\
127 & 622.19 & 143.00 & 18.06 & 16.46 & 16.03 &  0.11455 & -0.42732 \\
128 & 653.65 & 341.11 & 18.21 & 16.51 & 11.48 &  0.15582 &  0.35923 \\
129 & 435.61 & 237.91 & 18.00 & 16.35 &  8.70 & -0.14726 &  0.08191 \\
130 & 666.45 & 601.80 & 18.19 & 16.39 & 13.95 & -0.97424 & -0.14740 \\
131 & 373.24 & 178.94 & 18.15 & 16.31 & 11.57 &  0.41075 &  0.00232 \\
132 & 502.23 & 531.60 & 18.32 & 16.63 &  6.17 & -0.17885 &  0.14291 \\
133 & 101.71 & 572.66 & 18.18 & 16.53 & 16.08 &  0.27371 & -0.37341 \\
134 & 234.41 & 525.98 & 18.08 & 16.47 &  9.66 & -0.22959 & -0.24679 \\
135 & 527.26 & 464.69 & 18.36 & 16.65 &  5.28 &  0.03622 &  0.08226 \\
136 & 213.02 & 369.59 & 18.05 & 16.46 &  9.83 & -0.15828 &  0.03495 \\
137 & 250.01 & 539.40 & 18.13 & 16.54 &  9.36 & -0.45722 & -0.43293 \\
138 & 618.62 & 555.78 & 18.20 & 16.52 & 10.94 & -0.07833 & -0.19288 \\
139 & 599.73 & 231.76 & 18.75 & 16.44 & 12.21 &  0.34534 & -0.10718 \\
140 & 479.53 & 379.96 & 18.20 & 16.55 &  3.50 & -0.03636 & -0.13112 \\
141 & 397.68 & 634.78 & 18.98 & 16.51 &  9.64 &  0.24791 & -0.12747 \\
142 & 760.66 & 435.76 & 18.44 & 16.71 & 15.72 & -0.45511 & -0.64631 \\
143 & 245.66 & 347.20 & 18.19 & 16.57 &  8.77 & -0.21254 &  0.42830 \\
144 & 143.99 & 486.59 & 18.28 & 16.66 & 12.96 & -0.39742 & -0.34786 \\
145 & 497.53 & 735.73 & 18.56 & 16.57 & 14.69 &  0.20060 &  0.72802 \\
146 & 409.92 & 315.33 & 19.26 & 16.75 &  5.12 &  0.16649 &  0.14908 \\
147 & 343.81 & 790.84 & 18.59 & 16.83 & 17.13 &  0.92693 &  0.93157 \\
148 & 242.07 & 484.76 & 18.77 & 16.64 &  8.58 & -0.19366 & -0.10121 \\
149 & 731.05 & 433.17 & 18.52 & 16.80 & 14.35 & -0.08844 & -0.01713 \\
150 & 473.15 & 474.39 & 18.30 & 16.62 &  3.33 & -0.36948 &  0.17520 \\
151 & 222.98 & 408.93 & 18.27 & 16.68 &  9.06 & -0.08739 &  0.13982 \\
152 & 145.19 & 501.83 & 18.37 & 16.69 & 13.07 &  0.04099 & -0.16242 \\
153 & 351.87 & 471.22 & 18.30 & 16.37 &  3.72 & -0.01095 &  0.09335 \\
154 & 300.08 & 353.20 & 18.37 & 16.72 &  6.43 & -0.37032 &  0.22264 \\
155 & 573.05 & 265.03 & 19.03 & 17.04 & 10.25 &  0.14656 & -0.20524 \\
156 & 441.97 & 649.70 & 18.89 & 16.80 & 10.33 &  0.34913 &  0.35292 \\
157 & 419.03 & 397.64 & 18.67 & 17.04 &  1.31 & -0.34913 & -0.08423 \\
158 & 382.21 & 423.33 & 18.38 & 16.73 &  1.70 & -0.40009 &  0.19597 \\
159 & 491.39 & 518.76 & 18.57 & 16.94 &  5.40 &  0.01558 &  0.03594 \\
160 & 280.30 & 594.79 & 18.47 & 16.92 & 10.05 &  0.09981 &  0.35488 \\
161 & 403.15 & 195.38 & 18.59 & 16.89 & 10.64 &  0.03032 &  0.14122 \\
162 &  52.95 & 493.82 & 19.27 & 16.87 & 17.13 &  0.18579 & -0.79161 \\
163 & 554.81 & 427.33 & 19.39 & 16.89 &  6.24 &  0.42479 & -0.09518 \\
164 & 710.68 & 449.69 & 18.73 & 17.00 & 13.45 & -0.53597 & -0.25255 \\
165 & 229.32 & 382.84 & 18.83 & 16.86 &  8.96 & -0.30638 &  0.47056 \\
166 & 522.65 & 396.29 & 19.80 & 17.11 &  4.96 & -0.09012 & -0.24426 \\
167 & 294.38 & 237.72 & 19.00 & 16.83 & 10.40 &  0.18081 &  0.21022 \\
168 & 381.34 & 322.34 & 18.78 & 17.09 &  5.09 & -0.05334 &  0.19232 \\
169 & 616.79 & 586.90 & 19.84 & 16.99 & 11.72 & -0.30294 & -0.43546 \\
170 & 240.35 & 400.95 & 19.67 & 16.96 &  8.31 & -0.58511 & -0.22910 \\
171 & 270.03 & 453.71 & 19.98 & 17.29 &  6.98 & -0.84860 & -0.11722 \\
172 & 405.76 & 104.51 & 18.79 & 17.10 & 14.81 &  0.32849 & -0.75118 \\
173 & 608.04 & 127.54 & 18.91 & 17.16 & 16.26 &  0.10388 & -0.65688 \\
174 & 291.79 & 637.15 & 19.66 & 16.94 & 11.33 &  0.32568 & -0.01656 \\
175 & 574.51 & 366.38 & 19.78 & 17.05 &  7.66 &  0.17716 &  0.08240 \\
176 & 290.31 & 282.35 & 18.82 & 17.08 &  8.88 &  0.39152 &  0.10543 \\
177 & 377.34 & 446.03 & 18.43 & 16.77 &  2.13 &  0.01755 &  0.42648 \\
178 & 131.16 & 629.49 & 19.85 & 17.24 & 16.22 &  0.09335 &  0.13280 \\
179 & 362.71 & 704.77 & 19.06 & 17.33 & 13.07 &  0.52811 &  0.56854 \\
180 & 234.71 & 412.42 & 19.62 & 17.14 &  8.51 & -0.39335 &  0.12592 \\
181 & 381.51 & 415.86 & 18.60 & 16.92 &  1.80 & -0.24328 &  0.30449 \\
182 & 546.06 & 189.09 & 19.05 & 17.33 & 12.37 &  0.36948 & -0.29992 \\
183 & 726.07 & 416.65 & 19.32 & 17.53 & 14.13 &  0.06317 & -0.41623 \\
184 & 401.07 & 301.89 & 20.17 & 17.48 &  5.78 & -0.51843 &  0.17941 \\
185 & 383.01 & 623.72 & 19.68 & 17.29 &  9.24 &  0.25662 &  0.28160 \\
186 & 551.70 & 411.56 & 20.06 & 17.39 &  6.13 &  0.13336 &  0.07763 \\
187 & 577.22 & 744.61 & 19.67 & 17.45 & 16.35 & -0.75497 & -0.13027 \\
188 & 638.51 & 183.63 & 19.23 & 17.49 & 15.04 &  0.02583 & -0.05608 \\
189 & 540.50 & 694.19 & 19.90 & 17.56 & 13.53 &  0.27655 &  0.25240 \\
190 & 335.88 & 578.06 & 20.14 & 17.44 &  7.97 & -0.06219 & -0.21029 \\
191 & 143.77 & 267.34 & 19.93 & 17.40 & 14.62 & -0.03727 &  0.36008 \\
192 & 168.84 & 314.35 & 19.27 & 17.47 & 12.61 &  0.23921 &  0.09532 \\
193 & 283.33 & 574.14 & 19.71 & 17.71 &  9.24 &  0.71749 &  0.33860 \\
194 & 326.00 & 287.11 & 20.18 & 17.49 &  7.70 & -0.10529 &  0.36892 \\
195 & 100.00 & 403.04 & 19.35 & 17.47 & 14.72 &  0.28750 & -0.17014 \\
196 & 731.38 & 269.55 & 20.12 & 17.45 & 16.07 &  0.53962 &  0.16017 \\
197 & 680.15 & 535.04 & 20.45 & 17.66 & 13.01 & -0.25549 & -0.61318 \\
198 & 119.14 & 437.01 & 19.94 & 17.43 & 13.81 & -0.07268 & -0.36653 \\
199 & 497.98 & 232.01 & 19.42 & 17.68 &  9.64 &  0.15638 & -0.26686 \\
200 & 504.09 & 629.38 & 19.49 & 17.57 & 10.13 &  0.10851 & -0.04885 \\
201 & 506.81 & 430.19 & 20.45 & 17.68 &  4.04 & -0.77925 &  0.51071 \\
202 & 378.11 & 171.24 & 19.38 & 17.68 & 11.88 &  0.13673 & -0.07082 \\
203 & 256.40 & 594.80 & 20.37 & 17.68 & 10.78 &  0.21198 &  0.08591 \\
204 & 234.24 & 377.92 & 20.26 & 17.63 &  8.79 & -0.24054 & -0.15596 \\
205 & 101.74 & 352.14 & 19.58 & 17.70 & 14.99 & -0.10858 &  0.11090 \\
206 & 566.42 & 620.90 & 20.60 & 17.81 & 11.23 & -0.23724 & -0.13926 \\
207 & 689.64 & 369.47 & 19.47 & 17.91 & 12.71 &  0.32288 & -0.44304 \\
208 & 490.24 & 376.90 & 20.36 & 17.70 &  3.98 & -0.00730 &  0.15919 \\
209 & 340.77 & 530.73 & 20.58 & 18.03 &  6.01 & -0.50930 &  0.14431 \\
210 & 546.62 & 496.01 & 19.60 & 17.82 &  6.68 & -0.23640 & -0.20468 \\
211 & 117.87 & 638.09 & 20.97 & 18.03 & 16.94 &  0.11778 & -0.66035 \\
212 & 563.43 & 724.04 & 20.49 & 17.83 & 15.22 & -0.68955 &  0.03706 \\
213 & 159.79 & 308.04 & 20.61 & 17.88 & 13.11 &  0.19415 &  0.11188 \\
214 & 157.63 & 628.02 & 20.43 & 17.83 & 15.19 &  0.33993 & -0.58511 \\
215 & 512.09 & 179.66 & 19.89 & 18.07 & 12.12 &  0.49175 & -0.95354 \\
216 & 459.75 & 766.73 & 19.85 & 18.03 & 15.77 &  0.55156 &  0.57472 \\
217 & 675.83 & 519.51 & 20.96 & 18.14 & 12.56 & -0.73082 & -0.44248 \\
218 & 258.53 & 486.98 & 20.65 & 17.86 &  7.90 & -0.07637 & -0.22124 \\
219 & 313.93 & 596.28 & 20.39 & 17.92 &  9.20 &  0.22489 & -0.15835 \\
220 & 425.45 & 560.04 & 20.47 & 17.75 &  6.16 & -0.15877 & -0.07946 \\
221 & 204.88 & 370.95 & 20.66 & 17.91 & 10.18 & -0.09981 & -0.01011 \\
222 & 453.90 & 502.19 & 20.86 & 17.92 &  3.84 & -0.19134 & -0.33902 \\
223 & 382.37 & 528.90 & 19.50 & 17.24 &  5.02 & -0.00646 & -0.55029 \\
224 & 242.65 & 433.90 & 20.59 & 17.90 &  8.13 & -0.61683 & -0.20145 \\
225 & 402.07 & 287.19 & 20.95 & 18.19 &  6.44 &  0.12873 & -0.20608 \\
226 & 331.25 & 515.62 & 20.55 & 17.89 &  5.77 & -0.07216 &  0.04520 \\
227 & 547.83 & 500.57 & 20.87 & 18.09 &  6.84 &  0.03257 & -0.12564 \\
228 & 307.16 & 734.36 & 21.42 & 18.30 & 15.08 &  1.08192 &  0.43518 \\
229 & 363.68 & 184.93 & 21.15 & 18.21 & 11.39 & -0.07047 & -0.07082 \\
230 & 236.32 & 486.55 & 21.03 & 18.19 &  8.86 & -0.22215 & -0.11680 \\
231 & 422.46 & 586.79 & 20.79 & 18.04 &  7.39 & -0.05292 &  0.44248 \\
232 & 350.02 & 555.14 & 20.81 & 18.19 &  6.73 & -0.40107 & -0.17211 \\
233 &  95.01 & 358.38 & 20.90 & 18.07 & 15.23 & -0.07816 &  0.07272 \\
234 & 428.68 & 561.42 & 20.32 & 17.67 &  6.23 & -0.02850 & -0.07946 \\
235 & 641.32 & 161.10 & 20.95 & 18.27 & 15.91 &  0.16677 & -0.45266 \\
236 & 540.91 & 607.99 & 20.31 & 18.47 & 10.06 &  0.23893 & -0.03229 \\
237 & 466.77 & 272.93 & 21.14 & 18.33 &  7.38 &  0.32105 &  0.10346 \\
238 & 688.51 & 254.40 & 21.06 & 18.30 & 14.70 &  0.20917 & -0.38177 \\
239 & 706.83 & 337.01 & 21.36 & 18.54 & 13.85 &  0.16144 & -0.21899 \\
240 & 250.16 & 131.29 & 21.47 & 18.47 & 15.64 &  0.28883 & -0.28434 \\
241 & 423.77 & 182.03 & 20.30 & 18.42 & 11.23 & -0.02625 & -0.07314 \\
242 & 240.92 & 418.74 & 20.74 & 18.43 &  8.21 &  0.64259 & -0.58005 \\
243 & 127.84 & 489.44 & 20.63 & 18.49 & 13.71 & -0.27785 & -0.10753 \\
244 & 562.75 & 136.24 & 20.56 & 18.54 & 14.89 &  0.62469 & -0.75665 \\
245 & 431.40 & 730.28 & 20.54 & 18.71 & 14.00 &  0.46649 &  0.57472 \\
246 & 219.06 & 233.82 & 21.41 & 18.53 & 12.77 &  0.24658 &  0.18860 \\
247 & 426.46 & 587.68 & 20.64 & 18.05 &  7.43 & -0.41581 & -0.11427 \\
248 & 592.99 & 549.44 & 21.65 & 18.67 &  9.80 & -0.16565 & -0.11680 \\
249 & 407.14 & 554.03 & 20.67 & 17.86 &  5.91 &  0.32428 & -0.15358 \\
250 & 404.25 & 677.95 & 21.31 & 18.69 & 11.60 &  0.29480 &  0.22517 \\
251 & 225.81 & 373.98 & 21.63 & 18.80 &  9.21 & -0.15821 & -0.31866 \\
252 & 238.35 & 435.13 & 21.67 & 18.72 &  8.33 & -0.66702 &  0.14894 \\
253 & 403.79 & 155.25 & 20.59 & 18.67 & 12.48 & -0.46143 & -0.66976 \\
254 & 484.29 & 248.52 & 21.48 & 18.79 &  8.71 &  0.14572 &  0.14993 \\
255 &  90.01 & 592.62 & 21.71 & 18.82 & 16.97 & -0.41153 & -0.39559 \\
256 & 599.69 & 302.53 & 21.88 & 18.83 & 10.07 &  0.27908 & -0.10248 \\
257 & 186.43 & 496.98 & 22.11 & 18.99 & 11.19 & -0.26960 & -0.48614 \\
258 & 327.61 & 278.30 & 21.71 & 18.71 &  8.00 & -0.14038 &  0.36667 \\
259 & 197.21 & 450.92 & 22.06 & 19.01 & 10.27 & -0.17569 & -0.04787 \\
260 & 131.38 & 452.85 & 21.73 & 18.77 & 13.29 & -0.03587 & -0.19148 \\
261 & 183.51 & 507.34 & 21.60 & 18.79 & 11.46 & -0.30680 & -0.71019 \\
262 & 649.41 & 211.26 & 21.70 & 18.91 & 14.49 &  0.48319 & -0.11406 \\
263 & 486.86 & 418.10 & 21.52 & 18.46 &  3.14 & -0.21857 & -0.46747 \\
264 & 309.93 & 136.43 & 21.82 & 18.95 & 14.25 &  0.03173 & -0.29901 \\
265 & 531.46 & 724.53 & 21.10 & 19.07 & 14.66 &  0.22321 &  0.47786 \\
266 & 465.79 & 723.09 & 22.35 & 19.16 & 13.82 &  0.06710 & -0.12269 \\
267 & 470.67 & 620.37 & 21.77 & 18.98 &  9.24 &  0.12143 & -0.13870 \\
268 & 232.22 & 419.38 & 21.86 & 18.90 &  8.60 & -0.56672 &  0.25255 \\
269 &  87.98 & 542.33 & 21.84 & 18.70 & 16.14 &  0.47930 & -0.31417 \\
270 & 574.17 & 601.13 & 22.32 & 19.14 & 10.75 & -0.25325 &  0.07974 \\
271 & 559.22 & 530.71 & 21.81 & 19.02 &  8.04 &  0.03762 & -0.07665 \\
272 & 110.24 & 389.10 & 22.32 & 19.25 & 14.31 &  0.16425 & -0.07272 \\
273 & 706.85 & 345.66 & 21.59 & 19.04 & 13.74 & -0.12606 &  0.46227 \\
274 & 350.83 & 563.06 & 21.71 & 18.84 &  7.04 &  0.03720 & -0.29424 \\
275 & 371.11 & 309.50 & 21.86 & 18.89 &  5.81 & -0.03215 &  0.37201 \\
276 & 252.77 & 545.45 & 21.91 & 18.97 &  9.42 &  0.22587 & -0.30322 \\
277 & 294.07 & 344.15 & 21.96 & 19.04 &  6.88 & -0.22447 &  0.20060 \\
278 & 492.61 & 136.02 & 22.29 & 19.21 & 13.77 &  0.22629 & -0.86573 \\
279 & 504.27 & 244.87 & 22.08 & 19.08 &  9.22 &  0.03720 &  0.54145 \\
280 & 598.15 & 470.15 & 22.08 & 19.07 &  8.48 & -0.33579 & -0.49175 \\
281 & 535.65 & 471.43 & 21.73 & 18.96 &  5.75 & -0.35601 & -0.27318 \\
282 & 203.54 & 291.56 & 21.86 & 19.15 & 11.69 &  0.01614 &  0.45175 \\
283 & 130.39 & 548.45 & 22.21 & 19.33 & 14.42 &  0.07223 & -0.59802 \\
284 & 565.69 & 475.11 & 21.91 & 18.97 &  7.11 &  0.16425 &  0.11357 \\
285 & 374.71 & 193.74 & 21.93 & 19.27 & 10.89 &  0.28385 &  0.30870 \\
286 & 172.11 & 551.78 & 22.07 & 19.19 & 12.75 & -0.33347 & -0.74851 \\
287 & 382.03 & 311.09 & 22.31 & 19.22 &  5.56 &  0.58286 &  0.16060 \\
288 & 518.54 & 375.99 & 22.20 & 19.22 &  5.12 &  0.27374 &  0.21984 \\
289 & 542.15 & 430.54 & 21.91 & 18.95 &  5.66 & -0.17491 & -0.08647 \\
290 & 390.34 & 314.54 & 22.08 & 19.02 &  5.31 &  0.04141 &  0.08605 \\
291 & 312.83 & 713.58 & 22.26 & 19.44 & 14.09 &  0.43925 & -0.31698 \\
292 & 162.72 & 606.54 & 22.05 & 19.35 & 14.42 &  0.04548 & -0.99642 \\
293 & 185.59 & 676.73 & 21.53 & 19.02 & 15.76 &  0.30771 &  0.19541 \\
294 & 350.87 & 227.09 & 22.67 & 19.36 &  9.68 & -0.21212 &  0.06212 \\
295 & 575.71 & 373.76 & 22.37 & 19.33 &  7.60 &  0.22742 &  0.29073 \\
296 & 395.04 & 133.17 & 22.46 & 19.33 & 13.53 & -0.25212 & -0.68730 \\
297 & 331.37 & 730.53 & 22.62 & 19.49 & 14.57 &  0.95585 & -0.44220 \\
298 & 377.08 & 600.84 & 22.22 & 19.25 &  8.26 &  0.13168 &  0.12325 \\
299 & 377.58 & 676.63 & 22.36 & 19.44 & 11.68 &  0.62090 & -0.24482 \\
300 & 584.51 & 403.87 & 22.44 & 19.40 &  7.68 & -0.15807 & -0.19639 \\
301 & 369.68 & 336.49 & 22.29 & 19.37 &  4.71 & -0.21113 &  0.32007 \\
302 &  85.75 & 450.33 & 22.77 & 19.60 & 15.38 & -0.40416 & -0.46452 \\
303 & 573.63 & 350.44 & 22.03 & 19.33 &  7.91 & -0.02021 &  0.46410 \\
304 & 230.44 & 615.37 & 22.16 & 19.41 & 12.29 &  0.08788 & -0.67523 \\
305 & 157.54 & 461.24 & 22.30 & 19.32 & 12.14 & -0.39559 & -0.17393 \\
306 & 105.12 & 621.86 & 22.20 & 19.45 & 17.02 &  0.04829 & -0.93437 \\
307 & 385.25 & 392.83 & 21.92 & 19.03 &  2.19 & -0.31796 & -0.39559 \\
308 & 463.12 & 589.64 & 21.95 & 19.12 &  7.78 &  0.30084 &  0.30042 \\
309 & 255.82 & 417.31 & 22.68 & 19.57 &  7.52 & -0.21113 &  0.03734 \\
310 & 269.30 & 658.02 & 21.82 & 19.59 & 12.70 &  0.44978 &  0.07187 \\
311 & 451.33 & 725.92 & 21.57 & 19.71 & 13.87 &  0.19499 & -0.84341 \\
312 & 468.90 & 348.70 & 22.27 & 19.38 &  4.24 & -0.38549 & -0.24103 \\
313 & 352.97 & 565.98 & 22.49 & 19.49 &  7.11 & -0.51927 & -0.06851 \\
314 & 479.29 & 553.63 & 22.08 & 19.43 &  6.48 & -0.98309 & -0.22573 \\
315 & 225.82 & 307.63 & 22.50 & 19.59 & 10.43 &  0.46922 &  0.32105 \\
316 & 417.98 & 133.75 & 22.93 & 19.73 & 13.45 &  0.56588 & -0.68906 \\
317 & 491.65 & 221.39 & 23.02 & 19.71 &  9.99 & -0.10500 & -0.37257 \\
318 & 500.55 & 509.52 & 22.32 & 19.21 &  5.36 & -0.17660 & -0.73040 \\
319 & 310.01 & 634.66 & 22.49 & 19.46 & 10.82 &  0.61585 &  0.10472 \\
320 & 493.72 & 141.20 & 22.36 & 19.32 & 13.55 &  0.02766 & -0.74472 \\
321 & 328.22 & 643.91 & 22.77 & 19.86 & 10.85 &  0.61234 & -0.63284 \\
322 & 182.95 & 444.14 & 23.13 & 19.84 & 10.90 & -0.51562 & -0.01151 \\
323 &  63.96 & 479.98 & 22.77 & 19.71 & 16.52 &  0.29159 & -0.42872 \\
324 & 269.99 & 380.50 & 22.57 & 19.64 &  7.18 &  0.08647 &  0.39966 \\
325 & 286.80 & 493.20 & 22.65 & 19.66 &  6.82 & -0.16649 & -0.74921 \\
326 & 340.32 & 185.16 & 22.72 & 19.84 & 11.67 &  0.06078 &  0.37994 \\
327 & 571.79 & 391.68 & 22.30 & 19.57 &  7.20 & -0.49863 & -0.57303 \\
328 & 306.83 & 264.24 & 22.63 & 19.64 &  9.07 & -0.15681 &  0.17295 \\
329 & 458.58 & 345.75 & 22.36 & 19.63 &  4.12 &  0.16649 & -0.04927 \\
330 & 248.72 & 282.59 & 22.52 & 19.66 & 10.25 & -0.02709 &  0.68211 \\
331 & 216.86 & 587.50 & 22.50 & 19.70 & 11.90 &  0.16972 & -0.69573 \\
332 & 276.80 & 661.89 & 22.85 & 19.90 & 12.66 &  0.34492 &  0.28104 \\
333 & 493.62 & 148.09 & 22.66 & 19.71 & 13.24 & -0.03538 & -0.69832 \\
334 & 351.32 & 372.16 & 22.78 & 19.94 &  3.99 &  0.45638 & -0.25942 \\
335 & 314.21 & 634.73 & 22.86 & 19.78 & 10.74 &  0.12382 & -0.58174 \\
336 & 630.95 & 358.39 & 22.78 & 19.85 & 10.23 &  0.01320 & -0.24609 \\
337 & 518.27 & 514.03 & 22.64 & 19.80 &  6.09 &  0.76283 & -0.59185 \\
338 & 479.36 & 536.28 & 22.80 & 19.66 &  5.77 &  0.05012 & -0.43883 \\
339 & 197.39 & 541.10 & 22.90 & 19.89 & 11.49 &  0.30814 & -1.04050 \\
340 & 312.93 & 160.34 & 22.96 & 19.98 & 13.17 &  0.50832 & -0.33488 \\
341 & 630.16 & 230.87 & 22.80 & 19.83 & 13.23 & -0.85183 &  0.18629 \\
342 & 359.10 & 645.25 & 23.02 & 19.90 & 10.45 &  0.37075 &  0.24735 \\
343 & 430.92 & 257.46 & 22.71 & 19.83 &  7.78 &  0.01011 &  0.07777 \\
344 & 227.27 & 503.89 & 22.95 & 19.97 &  9.52 & -0.08781 &  0.13617 \\
345 & 494.82 & 542.73 & 23.13 & 19.93 &  6.39 &  0.08016 & -0.29115 \\
346 & 686.48 & 466.80 & 23.18 & 20.11 & 12.44 & -0.29396 & -0.46915 \\
347 & 494.75 & 616.96 & 23.02 & 20.04 &  9.44 & -0.09799 & -0.83162 \\
348 & 486.31 & 253.52 & 23.26 & 20.05 &  8.52 & -0.41216 &  0.13013 \\
349 & 513.33 & 224.84 & 23.14 & 20.13 & 10.23 & -0.24482 & -0.35558 \\
350 & 251.46 & 220.78 & 22.87 & 19.87 & 12.20 &  0.19646 &  0.01930 \\
351 & 491.54 & 196.42 & 22.88 & 19.76 & 11.08 &  0.23233 &  0.42409 \\
352 & 526.76 & 442.33 & 23.00 & 20.12 &  5.01 &  0.72380 & -0.85745 \\
353 & 184.48 & 235.34 & 23.22 & 20.33 & 13.91 & -0.03495 &  0.19366 \\
354 & 670.90 & 196.29 & 22.87 & 20.17 & 15.68 &  0.65417 &  0.04001 \\
355 & 479.29 & 101.17 & 23.28 & 20.06 & 15.21 & -0.04955 & -0.66976 \\
356 & 593.33 & 539.75 & 23.19 & 20.16 &  9.56 &  0.87935 & -0.29677 \\
357 & 135.32 & 454.56 & 22.91 & 19.94 & 13.12 & -0.92188 &  0.71580 \\
358 & 385.40 & 354.77 & 15.10 & 13.60 &  3.63 &  0.05419 &  0.22629 \\
359 & 440.93 & 405.71 & 11.92 & 10.32 &  1.38 & -0.13799 &  0.04590 \\
360 & 433.96 & 452.25 & 12.82 & 11.15 &  1.38 &  0.00000 &  0.32203 \\
361 & 335.29 & 483.77 & 12.86 & 11.34 &  4.68 & -0.50593 &  0.00000 \\
362 & 406.67 & 507.55 & 13.09 & 11.56 &  3.78 &  0.00000 & -0.18404 \\
363 & 423.50 & 492.46 & 13.12 & 11.61 &  3.05 &  0.09195 &  0.55198 \\
364 & 358.09 & 414.45 & 13.15 & 11.59 &  2.86 & -0.23008 &  0.69011 \\
365 & 411.37 & 509.35 & 13.32 & 11.80 &  3.84 &  0.00000 &  0.32203 \\
366 & 306.55 & 575.18 & 13.18 & 11.70 &  8.59 &  0.22994 &  0.00000 \\
367 & 430.67 & 443.18 & 13.22 & 11.66 &  0.94 &  0.13799 &  0.22994 \\
368 & 360.83 & 482.14 & 13.33 & 11.85 &  3.72 & -0.13799 & -0.04590 \\
369 & 424.64 & 481.43 & 13.46 & 11.90 &  2.55 &  0.09209 &  0.59802 \\
370 & 592.23 & 510.22 & 13.57 & 11.96 &  8.85 &  0.18390 & -0.18404 \\
371 & 508.60 & 581.52 & 13.62 & 12.02 &  8.24 &  0.78206 &  0.04604 \\
372 & 397.40 & 442.74 & 13.58 & 12.08 &  1.26 & -0.13799 & -0.23008 \\
373 & 456.37 & 371.75 & 13.66 & 12.08 &  3.03 & -0.13799 &  0.04604 \\
374 & 451.85 & 453.88 & 13.70 & 12.12 &  1.97 &  0.04604 &  0.23008 \\
375 & 328.58 & 495.70 & 13.93 & 12.46 &  5.25 & -0.32203 & -0.27599 \\
376 & 415.71 & 490.41 & 14.31 & 12.81 &  2.96 & -0.27599 & -0.04604 \\
377 & 519.94 & 282.52 & 13.81 & 12.26 &  8.07 &  0.00000 &  0.32189 \\
378 & 400.62 & 351.65 & 14.04 & 12.49 &  3.53 & -0.55198 & -0.18404 \\
379 & 458.24 & 569.57 & 13.97 & 12.46 &  6.83 &  0.27599 & -0.22994 \\
380 & 406.78 & 569.06 & 14.13 & 12.55 &  6.60 & -0.04604 & -0.45989 \\
381 & 395.27 & 497.29 & 14.22 & 12.74 &  3.45 &  0.00000 & -0.22994 \\
382 & 391.78 & 549.92 & 14.20 & 12.71 &  5.83 &  0.36794 & -0.09209 \\
383 & 512.30 & 419.52 & 14.40 & 12.81 &  4.30 & -0.09209 &  0.22994 \\
384 & 296.12 & 315.53 & 14.26 & 12.65 &  7.61 & -0.13799 &  0.87401 \\
385 & 402.49 & 382.63 & 14.18 & 12.66 &  2.15 &  0.32189 & -0.09195 \\
386 & 397.69 & 446.10 & 14.35 & 12.76 &  1.35 & -0.13799 &  0.18404 \\
387 & 264.53 & 464.97 & 14.23 & 12.73 &  7.33 & -0.82797 & -0.41398 \\
388 & 383.91 & 479.85 & 14.24 & 12.78 &  2.95 &  0.00000 &  0.09209 \\
389 & 342.85 & 107.87 & 14.29 & 12.73 & 15.06 &  0.32203 &  0.13803 \\
390 & 411.70 & 515.19 & 14.52 & 13.00 &  4.11 &  0.41412 &  0.50593 \\
391 & 752.70 & 234.66 & 14.32 & 12.80 & 17.69 & -0.13785 & -0.09195 \\
392 & 477.39 & 496.02 & 14.47 & 12.94 &  4.18 &  0.13813 & -0.78206 \\
393 & 336.64 & 433.13 & 14.96 & 13.46 &  3.81 & -0.09195 &  0.00000 \\
394 & 317.77 & 326.86 & 14.63 & 13.13 &  6.53 &  0.00000 & -0.04604 \\
395 & 343.07 & 484.10 & 14.85 & 13.34 &  4.40 & -0.18390 & -0.09195 \\
396 & 676.63 & 327.21 & 15.07 & 13.56 & 12.69 &  0.27599 &  0.22994 \\
397 & 302.45 & 467.14 & 14.61 & 13.07 &  5.69 & -0.09195 & -0.41398 \\
398 & 652.42 & 324.93 & 14.93 & 13.42 & 11.70 &  0.04604 & -0.32203 \\
399 & 318.17 & 548.02 & 14.77 & 13.18 &  7.28 &  0.27613 & -0.59802 \\
400 & 391.70 & 470.68 & 14.76 & 13.19 &  2.40 & -0.41398 & -0.27599 \\
401 & 547.33 & 266.18 & 14.74 & 13.21 &  9.43 & -0.36780 & -0.32203 \\
402 & 385.12 & 538.86 & 14.88 & 13.35 &  5.41 &  0.59802 & -0.55198 \\
403 &  93.93 & 382.69 & 14.83 & 13.19 & 15.09 &  0.00000 & -0.09195 \\
404 & 385.31 & 354.76 & 15.22 & 13.71 &  3.64 &  0.09195 &  0.13799 \\
405 & 631.82 & 261.16 & 14.84 & 13.35 & 12.38 & -0.04604 & -0.32203 \\
406 & 473.73 & 446.04 & 14.95 & 13.35 &  2.67 & -0.18390 & -0.04590 \\
407 & 525.52 & 360.33 & 14.89 & 13.35 &  5.76 &  0.18418 & -0.09209 \\
408 & 415.36 & 515.59 & 15.05 & 13.57 &  4.11 &  0.13799 &  0.32203 \\
409 & 339.21 & 346.45 & 14.95 & 13.38 &  5.19 &  0.09195 & -0.04590 \\
410 & 451.68 & 490.01 & 14.96 & 13.44 &  3.29 & -0.36808 & -0.04590 \\
411 & 396.13 & 537.73 & 15.10 & 13.60 &  5.24 & -0.04604 & -0.27599 \\
\enddata
\begin{flushleft}
$X$ and $Y$ are star positions in pixel coordinates on the PC chip
of WFPC2 based on the 2007 data. $r$ is the cluster-centric distance
in arcsecond. $\mu_x$ and $\mu_y$ are relative proper motions'
components along the $x$ and $y$ directions of (PC) pixel
coordinates (2007). $V$ and $I$ magnitudes are from the 1997 WFPC2
data.
\end{flushleft}
\end{deluxetable}

\begin{deluxetable}{ccccccccccccccc}[]
\tablecolumns{8} \tablewidth{0pc} \tablecaption{The $VI$ photometry
and relative proper motions of the member stars on the WFC2 chip of
WFPC2 (online Table 3).} \tablehead{ \colhead{No.} & \colhead{$X$} &
\colhead{$Y$} & \colhead{$V$} & \colhead{$I$} & \colhead{$r('')$} &
\colhead{$\mu_x$(mas/yr)} & \colhead{$\mu_y$(mas/yr)}} \startdata
 \hline
  1 & 550.24 & 101.76 & 16.91 & 15.33 & 41.30 & -0.08606 & -0.16502  \\
  2 & 719.24 &  79.45 & 17.27 & 15.20 & 52.46 & -0.16724 & -0.10483  \\
  3 & 568.64 & 183.71 & 16.93 & 15.37 & 47.71 & -0.38513 &  0.75699  \\
  4 & 306.83 &  94.90 & 18.51 & 16.65 & 28.67 & -0.27374 &  0.03670  \\
  5 & 570.07 & 323.19 & 18.77 & 17.02 & 57.72 & -0.17029 & -0.17395  \\
  6 & 612.66 & 166.41 & 18.55 & 16.89 & 49.38 & -0.61340 &  0.55496  \\
  7 & 435.92 & 216.66 & 18.86 & 17.19 & 43.29 & -0.39001 &  0.47302  \\
  8 & 650.87 & 108.70 & 19.07 & 16.99 & 48.71 & -0.26123 & -0.25101  \\
  9 & 541.53 & 200.57 & 18.80 & 17.20 & 47.29 &  0.02197 &  0.55008  \\
 10 & 598.51 & 207.28 & 19.05 & 17.26 & 51.09 &  0.18921 &  0.39795  \\
 11 & 499.89 &  58.45 & 19.65 & 17.27 & 35.33 & -0.27802 &  0.18848  \\
 12 & 434.89 &  60.65 & 19.10 & 17.42 & 31.46 &  0.00183 & -0.51270  \\
 13 & 756.71 & 133.74 & 19.76 & 17.41 & 57.86 & -0.00916 & -0.22888  \\
 14 & 393.03 & 183.24 & 19.55 & 17.71 & 38.89 & -0.01190 &  0.04791  \\
 15 & 380.83 & 193.30 & 20.50 & 17.65 & 39.28 & -0.12695 & -0.45609  \\
 16 & 618.54 & 158.25 & 19.45 & 17.79 & 49.27 &  0.16663 &  0.52109  \\
 17 & 505.67 & 257.21 & 20.14 & 18.02 & 49.58 &  0.06104 &  1.36597  \\
 18 & 557.61 & 125.01 & 19.97 & 18.08 & 43.22 & -0.33203 & -0.35904  \\
 19 & 598.44 & 292.96 & 20.93 & 18.08 & 56.99 &  0.12634 & -0.16205  \\
 20 & 437.48 & 138.01 & 21.01 & 18.23 & 37.23 &  0.65094 & -0.17105  \\
 21 & 377.79 & 105.99 & 21.01 & 18.38 & 32.04 &  0.60303 &  0.00603  \\
 22 & 424.70 & 137.99 & 20.55 & 18.65 & 36.62 &  0.30975 & -0.25406  \\
 23 & 674.67 & 177.47 & 21.41 & 18.66 & 54.21 &  0.02136 &  0.27100  \\
 24 & 681.05 & 245.23 & 21.08 & 19.02 & 58.73 & -0.41138 & -0.16800  \\
 25 & 604.91 & 318.79 & 21.07 & 18.93 & 59.20 &  0.90515 & -0.58594  \\
 26 & 323.33 &  77.77 & 20.63 & 18.77 & 27.70 & -0.48920 & -0.01053  \\
 27 & 447.54 & 136.91 & 21.77 & 18.87 & 37.65 &  0.25299 & -0.75409  \\
 28 & 550.40 & 156.52 & 20.66 & 18.88 & 44.80 & -0.88989 & -0.81909  \\
 29 & 354.72 & 155.30 & 21.80 & 19.02 & 35.23 & -0.15381 & -0.12100  \\
 30 & 700.93 & 157.17 & 21.03 & 18.92 & 54.93 & -0.15198 &  0.14694  \\
 31 & 577.11 & 280.32 & 20.84 & 19.18 & 54.93 & -1.06690 & -0.42877  \\
 32 & 330.58 &  53.92 & 20.89 & 19.09 & 25.96 & -0.11383 &  0.53989  \\
 33 & 509.13 & 184.44 & 22.09 & 19.11 & 44.35 &  0.09705 &  0.12909  \\
 34 & 597.13 & 280.49 & 21.44 & 19.17 & 56.03 & -0.52063 &  0.26398  \\
 35 & 633.55 & 263.40 & 21.21 & 19.03 & 56.95 &  0.27100 &  0.01221  \\
 36 & 605.45 & 234.53 & 21.67 & 19.42 & 53.33 & -0.11597 & -0.07095  \\
 37 & 360.84 &  84.45 & 22.25 & 19.40 & 29.61 &  0.30518 & -0.04715  \\
 38 & 552.52 & 245.27 & 21.80 & 19.57 & 51.08 & -0.27405 &  0.08804  \\
 39 & 661.16 & 257.19 & 21.98 & 19.76 & 58.23 & -0.18921 & -0.99121  \\
 40 & 337.96 &  98.91 & 22.62 & 19.77 & 29.95 &  0.10193 & -0.12527  \\
 41 & 553.96 & 235.75 & 22.80 & 19.75 & 50.47 &  0.87585 & -0.57800  \\
 42 & 312.83 &  92.39 & 22.73 & 19.78 & 28.62 &  0.09308 & -0.16891  \\
 43 & 475.60 &  66.45 & 22.58 & 19.74 & 34.28 &  0.67993 &  0.24353  \\
 44 & 747.63 & 165.59 & 22.27 & 19.88 & 58.77 &  0.02991 & -0.74402  \\
 45 & 405.26 & 153.35 & 21.73 & 19.63 & 36.96 & -1.03699 & -0.68298  \\
 46 & 733.53 &  55.20 & 22.58 & 19.97 & 52.55 &  0.01953 &  0.67150  \\
 47 & 575.66 & 235.96 & 22.08 & 19.96 & 51.70 & -0.54626 &  0.27206  \\
 48 & 525.22 &  64.82 & 22.96 & 19.94 & 37.40 & -0.23804 & -0.00397  \\
 49 & 403.81 & 141.45 & 23.17 & 20.14 & 35.95 &  0.52002 &  0.55801  \\
 50 & 369.42 &  57.80 & 22.98 & 20.08 & 27.87 & -0.45715 &  0.67631  \\
 51 & 630.77 &  59.81 & 22.28 & 20.00 & 44.73 & -0.23499 & -0.75268  \\
 52 & 349.77 & 139.17 & 22.87 & 20.17 & 33.71 & -0.41687 & -0.29800  \\
 53 & 340.80 & 137.12 & 23.20 & 20.16 & 33.25 &  0.78400 & -0.01892  \\
 54 & 326.88 &  61.06 & 22.59 & 19.75 & 26.42 & -0.40802 &  0.31590  \\
 55 & 567.24 & 118.30 & 23.10 & 20.14 & 43.43 &  0.64819 &  0.08797  \\
 56 & 614.13 & 201.20 & 22.43 & 20.12 & 51.67 &  1.14258 & -0.29999  \\
 57 & 741.31 &  83.13 & 22.14 & 20.14 & 54.35 & -0.31433 & -0.14107  \\
 58 & 424.91 &  55.56 & 23.41 & 20.32 & 30.54 & -0.40100 & -0.37369  \\
 59 & 466.82 & 133.40 & 23.64 & 20.27 & 38.39 &  0.97809 & -0.24704  \\
 60 & 371.79 &  92.15 & 23.37 & 20.22 & 30.68 &  0.30304 & -0.01244  \\
 61 & 492.06 &  97.91 & 22.82 & 20.27 & 37.36 &  1.08002 & -0.36453  \\
 62 & 493.20 & 233.06 & 23.26 & 19.88 & 47.14 & -0.70984 & -0.05203  \\
 63 & 645.85 & 115.29 & 23.65 & 20.52 & 48.70 &  0.49988 &  0.19806  \\
 64 & 638.54 & 197.21 & 22.99 & 20.58 & 52.98 & -0.16724 & -0.45792  \\
 65 & 512.09 & 189.55 & 22.77 & 20.74 & 44.88 & -0.47913 & -0.62897  \\
 66 & 463.45 & 245.64 & 23.47 & 20.86 & 46.76 &  1.23108 &  0.32104  \\
 67 & 561.30 &  81.19 & 23.43 & 21.01 & 40.83 &  0.59326 & -0.49355  \\
 68 & 695.94 &  94.14 & 23.45 & 21.14 & 51.34 &  0.00488 &  0.48660  \\
\enddata
\begin{flushleft}
$X$ and $Y$ are star positions in pixel coordinates on the PC chip
of WFPC2 based on the 2007 data. $r$ is the cluster-centric distance
in arcsecond. $\mu_x$ and $\mu_y$ are relative proper motions'
components along the $x$ and $y$ directions of (PC) pixel
coordinates (2007). $V$ and $I$ magnitudes are from the 1997 WFPC2
data.
\end{flushleft}
\end{deluxetable}

\begin{deluxetable}{ccccccccccccccc}[]
\tablecolumns{8} \tablewidth{0pc} \tablecaption{The $VI$ photometry
and relative proper motions of the member stars on the WFC4 chip of
WFPC2 (online Table 4).} \tablehead{ \colhead{No.} & \colhead{$X$} &
\colhead{$Y$} & \colhead{$V$} & \colhead{$I$} & \colhead{$r('')$} &
\colhead{$\mu_x$(mas/yr)} & \colhead{$\mu_y$(mas/yr)}} \startdata
 \hline
  1 & 331.16 & 211.81 & 17.89 & 14.87 & 48.98 & -0.07019 &  0.22003  \\
  2 & 194.24 & 169.76 & 17.31 & 15.33 & 36.84 & -0.16998 &  0.34409  \\
  3 & 121.32 & 188.19 & 18.00 & 16.17 & 30.16 & -0.44304 &  0.15900  \\
  4 & 331.39 & 242.91 & 18.56 & 16.51 & 49.10 & -0.24994 &  0.64194  \\
  5 & 362.65 & 200.63 & 18.88 & 16.99 & 51.83 &  0.24292 &  0.54993  \\
  6 & 387.38 & 176.65 & 19.17 & 17.10 & 54.13 & -0.12604 &  0.32410  \\
  7 & 188.60 &  71.17 & 18.86 & 17.32 & 38.24 &  0.39307 &  1.06903  \\
  8 & 317.60 & 345.95 & 19.42 & 17.26 & 49.32 &  0.57281 & -0.01221  \\
  9 & 177.99 & 120.84 & 19.36 & 17.32 & 36.09 & -0.55801 &  0.50400  \\
 10 &  69.27 &  78.94 & 19.48 & 17.46 & 28.00 & -0.85289 & -0.33165  \\
 11 & 165.81 & 225.45 & 19.40 & 17.50 & 34.13 & -0.07401 &  0.05096  \\
 12 & 286.92 &  99.31 & 19.80 & 17.42 & 46.08 & -0.10284 &  0.15472  \\
 13 & 446.87 & 215.81 & 19.75 & 17.62 & 59.40 & -0.41992 &  0.69992  \\
 14 & 352.35 & 242.43 & 19.68 & 17.73 & 50.98 & -1.16913 & -0.14999  \\
 15 & 354.09 & 245.77 & 20.15 & 18.22 & 51.15 &  0.03998 & -0.06592  \\
 16 & 285.43 & 299.33 & 20.23 & 18.15 & 45.59 &  0.15808 & -0.47211  \\
 17 &  58.98 & 111.96 & 20.76 & 18.39 & 26.02 & -0.23930 & -0.26199  \\
 18 &  86.85 &  93.25 & 20.39 & 18.56 & 28.96 & -0.78171 &  0.26367  \\
 19 & 158.24 & 100.59 & 21.44 & 18.77 & 34.83 &  0.14709 & -0.45296  \\
 20 & 440.08 & 121.06 & 20.97 & 18.85 & 59.32 & -0.16296 & -0.30693  \\
 21 &  80.86 & 121.03 & 22.24 & 19.26 & 27.63 &  0.60005 & -0.37399  \\
 22 & 377.99 &  92.63 & 21.35 & 19.25 & 54.23 & -0.18524 & -0.54657  \\
 23 & 226.03 & 153.90 & 21.07 & 19.28 & 39.84 & -0.01404 &  0.52094  \\
 24 & 417.73 & 272.87 & 21.20 & 19.43 & 57.06 &  0.12390 & -0.32196  \\
 25 & 267.39 & 275.64 & 22.75 & 19.60 & 43.65 &  0.21118 & -0.44678  \\
 26 & 109.85 & 151.71 & 21.35 & 19.39 & 29.53 &  0.12604 & -0.25513  \\
 27 & 250.84 & 135.20 & 21.49 & 19.45 & 42.29 & -0.10712 & -1.21704  \\
 28 & 226.51 &  76.14 & 22.13 & 19.73 & 41.35 & -0.29190 & -0.36797  \\
 29 & 386.56 & 203.95 & 22.28 & 20.12 & 53.97 &  0.29297 & -1.01608  \\
 30 & 155.64 & 196.91 & 22.01 & 19.95 & 33.21 &  0.25696 & -1.33789  \\
 31 & 397.20 & 142.06 & 22.92 & 20.25 & 55.26 &  1.11816 & -0.46707  \\
 32 & 394.49 & 392.47 & 23.07 & 20.28 & 57.11 & -0.54321 &  0.22980  \\
 33 & 288.93 & 111.01 & 23.02 & 20.45 & 46.05 &  0.16693 & -1.11801  \\
\enddata
\begin{flushleft}
$X$ and $Y$ are star positions in pixel coordinates on the PC chip
of WFPC2 based on the 2007 data. $r$ is the cluster-centric distance
in arcsecond. $\mu_x$ and $\mu_y$ are relative proper motions'
components along the $x$ and $y$ directions of (PC) pixel
coordinates (2007). $V$ and $I$ magnitudes are from the 1997 WFPC2
data.
\end{flushleft}
\end{deluxetable}